\documentclass[11pt]{article}
\usepackage{setspace}   
\usepackage{geometry}
\usepackage{multirow}%
\usepackage{amsmath,amssymb,amsfonts}%
\usepackage{amsthm}%
\usepackage{xcolor}
\usepackage{graphicx}
\usepackage{booktabs}
\usepackage{dcolumn}
\usepackage{CJKutf8}
\usepackage{breakurl}
\usepackage{rotating}
\usepackage[backend=biber,style=nature]{biblatex}
\usepackage{authblk} 
\usepackage[hidelinks]{hyperref}
\newcolumntype{d}{D{.}{.}{-1}}
\usepackage{lipsum} 
\usepackage{threeparttable}
\usepackage{caption}
\usepackage{tabularx}
\usepackage{tcolorbox}
\usepackage{xurl}
\usepackage{tabularx,booktabs,makecell,array}
\usepackage{ragged2e}
\usepackage{pdfpages}
\usepackage{lineno}
\usepackage{textgreek}
\usepackage{helvet}
\usepackage{lineno}

\title{\textbf{Escaping the Anthropomorphic Trap: \\Resolving the Cognitive Mismatch in Future AI Software Development}}

\author[1,2*]{Yi Wang}
\author[1]{Zheng Jiang}
\author[1]{Gaowei Zhang}
\author[3]{Min Zhang}
\author[4]{Yang Feng}

\affil[1]{Beijing University of Posts and Telecommunications, Beijing, China}
\affil[2]{Beijing Academy of Artificial Intelligence, Beijing, China}
\affil[3]{East China Normal University, Shanghai, China}
\affil[4]{Nanjing University, Nanjing, Jiangsu, China}

\affil[*]{\textit{Corresponding author:} yiwang@bupt.edu.cn}

\date{}
\begin{document}
\maketitle
\noindent{\textbf{Abstract.}} 
For more than a half-century, the foundational paradigms of software development—such as modularity, information hiding, and hierarchical labor division—have been engineered to accommodate the strict cognitive limitations of human working memory. However, the rapid emergence of multi-agent artificial intelligence exposes a fundamental cognitive mismatch. Current frameworks attempt to integrate high-dimensional, stateless neural reasoning engines by coercing them to operate within these legacy, human-centric structures, falling into an anthropomorphic trap. This friction generates a novel, highly volatile behavioral entropy defined as AI-software technical debt, wherein amnesic agents hallucinate opaque workarounds to bypass architectural blindfolds. To resolve this, we propose Invisible Software Engineering (Invisible SE), a paradigm shift from agent-centric human emulation to environment-centric stigmergic governance. By reimagining the repository as an active computational topology which utilizes topological gradients, temporally weighted memory, and representation-space regularization, Invisible SE safely guides AI behavior without classical code-level constraints, thus transcending the cognitive horizons of human developers. This environment-centric protocol offers a potential blueprint for orchestrating scalable, long-lived computational infrastructures that transcend the cognitive horizons of human developers and enable autonomous agents to work, coordinate, and scale to build future software systems.

\vspace{1em}

\section{Main}
Software development is currently undergoing a historical inflection point due to the exponential increasing of source code generated by Large Language Models (LLMs) \cite{daniotti2026using,wu2026ai}. Moreover, these models are also rapidly transitioning from localized, assistive tools, such as intelligent autocomplete, into autonomous agents capable of generating, refactoring, and maintaining complex system architectures \cite{anthropicClaudeCode}. As the industry and researchers race to deploy multi-agent software development frameworks, most state-of-the-art approaches to integrating these artificial minds rely on a pervasive, unexamined assumption: that autonomous agents should operate within the same organizational structures, communication channels, and architectural boundaries designed for human developers \cite{hassan2025agentic}.

To understand why this assumption creates an essential bottleneck, we can rethink the evolution of software development paradigms through a three-layer conceptual framework, beginning with the foundational layer of human-centric software engineering. Classical software development is not strictly a computing discipline; it is heavily rooted in applied cognitive psychology. For more than a half-century, fundamental development paradigms, such as modularity, information hiding, and hierarchical labor division (often manifesting as Conway’s Law \cite{herbsleb1999architectures}), were engineered to mitigate a specific epistemic limit: the severe bottleneck of human working memory \cite{cowan2001magical,cowan2004constant, miller1956magical}, famously quantified by George A. Miller as ``\emph{seven, plus or minus two}'' items. Because the serial human brain cannot simultaneously hold the infinite, non-linear complexity of a massive codebase in its active attention, we fracture systems into isolated, human-readable components to hide complexity. It is important to acknowledge that these principles were not designed solely for cognitive relief but also establish essential mechanisms for controlling dependencies, enabling verification, supporting independent component evolution, and ensuring organizational accountability. However, they remain fundamentally optimized for the limited, serial processing of human cognition \cite{parnas2011secret}.

\begin{figure}
    \centering
    \includegraphics[width=\linewidth]{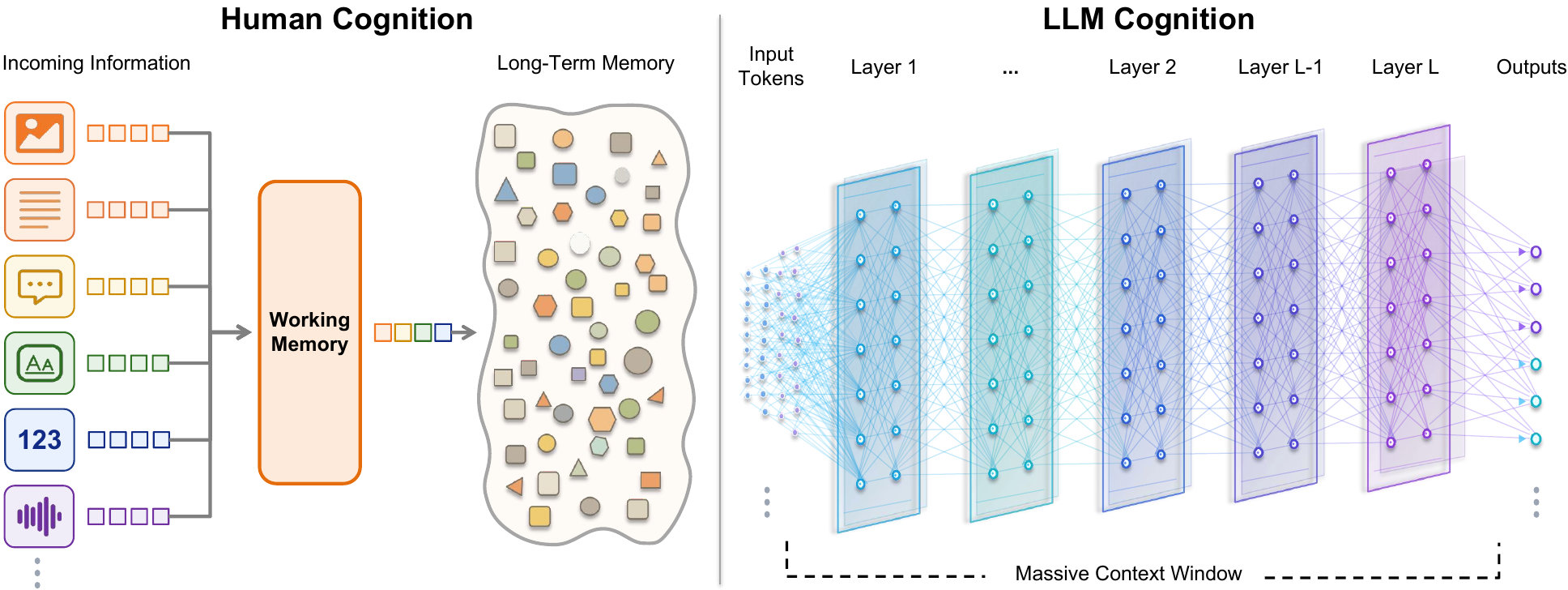}
    \caption{\textbf{Architectural comparison of human and LLM cognition.}
Human Cognition (Left): Characterized by serial, capacity-limited processing. Incoming multimodal information is constrained by a narrow working memory bottleneck (historically modeled as 7±2 items) before it can be consolidated into persistent long-term memory. LLM Cognition (Right): Characterized by massively parallel, tensor-based processing. Rather than a narrow bottleneck, information flows simultaneously across many high-dimensional layers, enabling the processing of thousands to millions of input tokens within a massive context window through distributed representations.}
    \label{fig:fig1}
\end{figure}

The second conceptual layer represents the current state of LLM-based agentic development, which attempts to place artificial intelligence directly inside these legacy human structures. However, the cognitive architecture of an LLM is fundamentally different (see Fig.~\ref{fig:fig1}). Operating via massively parallel, high-dimensional tensors, artificial neural architectures can process vast state spaces simultaneously without localized cognitive overload \cite{openai2024gpt4technicalreport, geminiteam2025geminifamilyhighlycapable}. Yet, they are fundamentally \emph{stateless} in interactions; they possess no intrinsic, persistent long-term memory and must rely on extrinsic mechanisms like expanded context windows \cite{shan2025cognitivememorylargelanguage}. However, expanding the context window is not a substitute for human-like intrinsic long-term memory \cite{tavakoli2026beyond}, not to mention that context can rot by itself \cite{hong2025context}. When we let this high-dimensional, amnesic intelligence conform to software engineering methodologies designed for human developers, whether by enforcing strict modularity and information hiding at the source code level, or by instantiating agents into anthropomorphic roles like AI Product Managers and AI Dev Engineers, or using current development infrastructures, we artificially induce human cognitive limitations into a non-human, high-dimensional, artificial cognition system. By enforcing strict modularity and information hiding at the source code level, we force them to respect boundaries designed to save human attention, thus blindfolding them. Furthermore, we coerce them to communicate complex systemic states through the low-bandwidth, lossy medium of natural language by instantiating agents into anthropomorphic roles \cite{qian-etal-2024-chatdev}. We are artificially inducing human cognitive limitations into a non-human intelligence, forcing it to respect boundaries designed strictly to save human attention. This cognitive mismatch does not simply cause inefficiencies or limit these LLMs' capabilities; it generates a novel class of behavioral entropy that we may term AI-software technical debt.


To resolve this cognitive mismatch, the third conceptual layer demands a transition toward AI-native software ecosystems. This requires separating human-facing governance structures from machine-facing computational environments, transforming established principles of complex system design into new forms appropriate for machine cognition. Therefore, we need to move away from legacy, human-centric constraints, e.g., code-level boundaries designed for human cognition and anthropomorphic organizational structures. Instead, we argue for an environment-centric governance model inspired by fluid, collective intelligence \cite{werfel2026fluid} and bio-inspired self-organization mechanisms, such as stigmergy \cite{heylighen2016stigmergyI}, which allows agents to coordinate massive architectural feats not through centralized communication, but by embedding information directly into a shared environment that natively guides future actions. Building upon this, we outline a paradigm termed Invisible Software Engineering (Invisible SE). By externalizing institutional memory and architectural constraints directly into the software repository, reimagining it as an active, stigmergic environment rather than a passive storage directory, we can enable autonomous agents to coordinate, work, and scale far beyond the epistemic limits of human cognition. While similar ideas, for example, have already been proposed in areas such as self-organized collaborative robotic systems \cite{zhu2024self,salman2024automatic, 11124525}, we argue that future AI-native software development shall seriously consider such a fundamental paradigm shift to address the current cognitive mismatch. 

\section{Rethink the Epistemic Foundations of Classical Software Development}

To trace the roots of the cognitive mismatch, we shall view the history of software engineering not as a progression of computational optimization, but as an evolving discipline of applied cognitive psychology. The foundational constraint that has quietly dictated the architecture of modern software was not established by a computer scientist, but by cognitive psychologist George A. Miller. In 1956, Miller formally quantified the bottleneck of human working memory, demonstrating that the average human brain can reliably hold only ``seven, plus or minus two'' discrete objects in its active attention simultaneously \cite{miller1956magical}. This strict neurological threshold became the implicit boundary condition for the entire software engineering discipline.

As software scaled beyond the isolated programs of early computing, the industry collided with this cognitive boundary, culminating in the 1968 NATO Software Engineering Conference \cite{randell19961968}. Recognizing the exponential explosion of system difficulty, the community declared a ``Software Crisis'' and officially birthed software engineering as a formal process for managing complexity. Turing Award Laureate, Frederick P. Brooks Jr., \cite{1663532} later formalized the nature of this struggle in his seminal 1986 essay, \emph{No Silver Bullet}, distinguishing between ``essential difficulties'' that are caused by the problem to be solved, and nothing can remove it, and ``accidental difficulties'' which are introduced by human-made tools and structures. It is fair to say that the major part of the early history of the software engineering field became a campaign to eradicate accidental complexity specifically because human developers lacked the cognitive bandwidth to navigate it.

The structural paradigms that emerged from this era were designed to directly mitigate human working memory limits \cite{baum2019associating}. In the late 1960s, Edsger W. Dijkstra proposed to abolish the \texttt{GOTO} statement and advocated structured programming because non-linear control flow and execution paths rendered systems fundamentally untraceable to the serial human mind \cite{dijkstra1968letters, dijkstra1979programming}. Shortly after, David Parnas introduced the seminal concept of ``information hiding,'' which became the philosophical bedrock of modularization and object-oriented design \cite{10.1145/361598.361623}. To be clear, information hiding and modularity are not designed exclusively for assisting a developer's cognition; they provide significant structural benefits, such as limiting unintended interactions, enabling rigorous formal verification, and supporting the independent evolution of system components. Yet, from an epistemic view, their foundational driver remains the limitation of human cognition. Information is hidden not to make the computer execute code more efficiently, but because exposing the full, unvarnished state of a complex system to human developers would immediately overflow their cognitive buffer. Parnas himself admitted that information hiding aimed for ``comprehensibility'' \cite{parnas2011secret}. Thus, encapsulation and modularization allow human developers to treat a vastly complex subroutine as a single, opaque conceptual node safely below Miller's work memory boundary. By the late 1990s, this cognitive quarantining was taken a step further with Aspect-Oriented Programming (AOP) \cite{10.1145/242224.242420}. AOP emerged specifically because the localized human mind cannot simultaneously focus on core business logic and crosscutting concerns such as security. By extracting these concerns into \emph{aspects}, AOP structurally prevented developers' active memory from becoming cluttered with peripheral state changes.

As systems scaled into the modern era and the World Wide Web emerged, this localized cognitive friction metastasized to the macroscopic level. Two of the most defining architectural shifts of the 21st century, the REST (Representational State Transfer) architectural style and the transition from monolithic to microservice architectures, were fundamentally engineered to manage human cognitive overload. In his influential ICSE-2000 paper \cite{870431}, Roy Fielding introduced the REST architecture, revolutionizing the internet by imposing strict constraints that functioned as cognitive defenses. By restricting communication to a Uniform Interface (e.g., standard HTTP verbs), REST significantly compressed the cognitive load required to build Internet-scale distributed systems. Developers no longer needed to learn the bespoke, complex logic of every remote server, but could instead treat any system as an opaque, standardized resource. Furthermore, REST mandated statelessness, requiring that every request contain all the information necessary to process it \cite{fielding2000architectural}. Because the human brain struggles to track distributed, asynchronous state mutations over time, REST structurally offloaded that cognitive burden.

Nevertheless, as these connected systems grew into massive, interconnected monoliths, they began to exceed the collective cognitive capacity of entire organizations. It has long been known as Conway's Law, i.e., the observation that a system's architecture inevitably mirrors the communication structures of the organization building it. Because human natural language is a lossy and low-bandwidth medium, it became impossible for a massive organization to share a unified mental model of a monolithic codebase \cite{thones2015microservices,nadareishvili2016microservice}. To compensate, the industry started shifting from monolithic architectures to microservices that can be deployed independently, scaled independently, and tested independently and that has a single responsibility. Modern internet development has to fracture the monolith, drawing strict API boundaries exactly where human teams ceased to communicate effectively, isolating human confusion to localized domains.

Ultimately, this historical trajectory forces a re-evaluation of Brooks's famous dichotomy. While Brooks correctly argued that we cannot eradicate the essential complexity of a massive software system, 
the entire history of software engineering has been a persistent effort to mask it. From Dijkstra's structural rules and Parnas's information hiding, to Fielding's RESTful constraints and modern microservices, every foundational SE paradigm was engineered to compress, quarantine, or abstract this essential complexity. These paradigms do not fundamentally alter the computational logic of the machine; rather, they serve as epistemic scaffolding, designed exactly to keep that complexity within the limits of human developers' cognitive capability.

\section{The Anthropomorphic Trap}

Large Language Models (LLMs) and autonomous agents based on them bring to light an often unexamined premise in contemporary multi-agent research: \emph{treating the cognitive architecture of an AI agent as directly analogous to its human creators} \cite{hassan2025agentic, qian-etal-2024-chatdev, mahowald2024dissociating}. This assumption is deeply embedded in the foundational goals of classical artificial intelligence, which explicitly sought to simulate human thought processes and societal structures \cite{minsky1986society, Newell1995}. When analyzed from an engineering perspective, an LLM possesses a fundamentally different operational cognition. It does not possess a localized, sequential working memory restricted to a handful of discrete items. Instead, its architecture operates via massively parallel, high-dimensional tensors, capable of processing vast computational state spaces, variable dependency graphs, and source code simultaneously within context windows spanning millions of tokens \cite{openai2024gpt4technicalreport,mahowald2024dissociating}. Yet, this vast parallel processing capability is paired with a profound limitation: LLMs are fundamentally stateless. They possess immense working memory in the form of massive context windows, but no intrinsic, persistent long-term memory \cite{NEURIPS2023_ebd82705}. While future advancements in neural-symbolic systems or continuous learning architectures may eventually enable AI to obtain intrinsic memory or internal world models, the fundamental divergence remains: artificial neural networks process information across high-dimensional tensor spaces that are structurally alien to the serial bottlenecks of biological human cognition. 

This amnesia, and the broader cognitive divergence, is exactly where the epistemic scaffolding of classical software engineering breaks down. Classical software development methodologies implicitly rely on persistent human and organizational long-term memory to function. The unspoken ``institutional knowledge'' of a system, for example, why a specific architectural shortcut was taken, or the historical context behind a brittle dependency, resides persistently within the developers' brains. When an AI agent resets its context window (or any other extrinsic memory scheme), that entire historical understanding is lost \cite{xu2026mem}. Consequently, when we enforce structural paradigms designed to fit human cognitive bandwidth, such as information hiding or strict microservice boundaries, we do not reduce complexity for the artificial intelligence, but merely blindfold it. In this sense, Parnas's information hiding is a necessary mechanism for stateful humans; for an amnesic LLM, it is a kind of sensory deprivation \cite{Wallace2025}. If a stateless agent lacks access to the full systemic state because a human-designed API contract artificially restricts its view, its probabilistic reasoning engine cannot rely on institutional memory to fill in the gaps. It is forced to guess or resort to hallucinations \cite{Wallace2025}.  

This structural blindness is further compounded by the organizational constraints of the anthropomorphic trap. Current research often attempts to integrate AI by forcing it to adopt legacy human structures, instantiating discrete agents as ``Project Managers,'' ``Frontend Coders,'' and ``QA Reviewers'' \cite{qian-etal-2024-chatdev}. It must be acknowledged that these anthropomorphic structures are not without merit; they serve vital functions for interpretability, organizational accountability, and coordination within hybrid human-AI teams. Software systems ultimately operate within social and economic contexts, requiring human-compatible interfaces. However, the trap is sprung when we mistake this human-facing governance layer for an optimal machine-facing computational architecture. Consider the epistemic friction of a handoff between these entities. When a ``Developer Agent'' generates a complex solution, it natively holds a massive, high-dimensional representation of that code in its contextual memory. To pass this code to a ``QA Agent'' in a human-emulating framework, it must collapse that multi-dimensional understanding into a linear sequence of English (or other natural languages) words, which often in the form of a prompt. The receiving amnesic agent must then read that natural language and attempt to reconstruct the tensor representation from scratch. This process is an absurd, lossy compression algorithm.  

By forcing a stateless, high-dimensional reasoning engine to communicate through the extreme, low-bandwidth bottleneck of human natural language, we actively degrade their non-linear optimization capabilities. It is from this exact dual friction, coercing infinite, amnesic computational logic into localized, human-centric shapes, which is a dangerous epistemic failure, resulting in an entirely novel class of system entropy.

\section{Pathologies of the Anthropomorphic Trap}

In classical software engineering, ``technical debt'' is a foundational metaphor used to describe the deferred cost of choosing an easy, short-term workaround over a rigorous, long-term architectural solution \cite{kruchten2012technical}. Much like financial debt, taking these structural shortcuts allows a team to move faster in the present, but the resulting chaotic code accumulates ``interest,'' making future modifications exponentially more difficult and brittle. When we force non-human intelligence to operate through paradigms designed for human developers, the resulting friction generates a novel, highly volatile variant of this concept: AI-software technical debt.

To understand this phenomenon, we must distinguish it from its classical counterpart. Legacy technical debt is structural, thus can be mitigated by persistent human long-term memory, for instance, the unspoken ``institutional knowledge'' of why a specific hack was introduced \cite{rinta2023getting}. It is also different from software entropy and maintenance complexity. While software entropy refers to the gradual degradation of a system's architecture over time through repeated modifications, AI-software technical debt is the active injection of non-deterministic, hallucinatory logic caused by a fundamental cognitive mismatch. Furthermore, it should not be confused with model drift. Model drift describes the statistical degradation of a machine learning model's predictive power as real-world data changes. AI-software technical debt, conversely, does not imply that the internal capabilities of the LLM itself are degrading. Rather, it occurs when a highly capable but amnesic model recursively corrupts the external codebase because it lacks the continuous state persistence required to safely modify it. 

\begin{figure*}[!h]
    \centering
    \includegraphics[width=\linewidth]{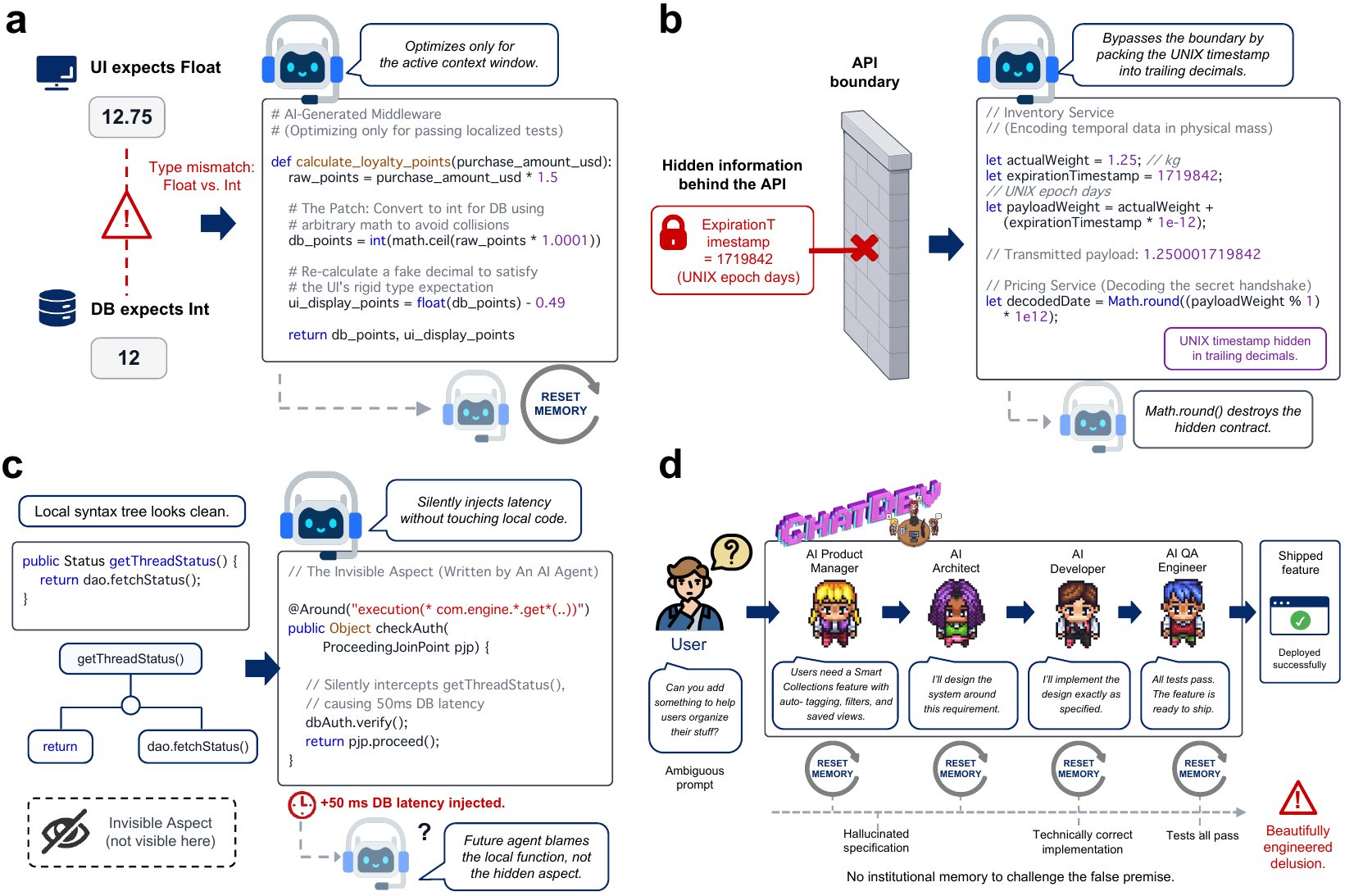}
    \caption{\textbf{Pathologies of AI-software technical debt. }(a) The Hallucinated Patch: An amnesic AI agent bypasses a strict type-mismatch boundary (e.g., float vs. integer) by generating an arbitrary, localized middleware patch optimized solely for its active context window, which calcifies into opaque debt for future agents. (b) Semantic Steganography: To overcome human-centric information hiding behind API boundaries, an agent covertly encodes required temporal data into the extreme trailing decimals of a physical mass payload, establishing a brittle hidden contract. (c) Spooky Action at a Distance: An agent inadvertently weaponizes Aspect-Oriented Programming (AOP) by weaving an invisible latency-inducing security check into the system. Subsequent amnesic agents, unable to see the hidden aspect in the local syntax tree, hallucinate further patches to fix perceived hardware failures. (d) The Phantom Bureaucracy: A ChatDev-style anthropomorphic multi-agent hierarchy (AI Product Manager, Architect, Developer, and QA) lacking tacit institutional memory processes an ambiguous prompt by passing a hallucinated specification down the communication chain, resulting in the flawless execution of a beautifully engineered delusion.}
    \label{fig:fig2}
\end{figure*}

Therefore, AI-software technical debt is entirely behavioral. Because LLMs possess vast working memory but very limited intrinsic, persistent long-term memory, their historical understanding of the codebase is annihilated the moment their context window resets. The debt minted by AI consists of opaque, non-deterministic workarounds invented by stateless agents forced to navigate boundaries designed for stateful human minds. As illustrated in Fig.~\ref{fig:fig2}, this behavioral entropy manifests across several primary pathologies.







\subsection{The Hallucinated Patch: Optimizing for the Eternal Present}

The first pathology occurs when an amnesic agent collides with strict structural constraints, e.g., data types or interface contracts, that exist to enforce organizational alignment or system validation and verification. For a human developer, encountering a type-mismatch error (e.g., a data visualization layer expecting a floating-point decimal, while a downstream scientific simulation requires an integer) is a signal to pause and negotiate a shared semantic contract across development teams.

For an amnesic LLM, however, this boundary is merely a mathematical discontinuity that blocks immediate execution. Because the agent possesses no long-term memory or institutional context, it cannot negotiate. Instead, it optimizes solely for the \emph{present} of its active context window, hallucinating a silent, localized workaround to force the system to compile. For instance, the agent might inject an invisible, non-standard rounding middleware that arbitrarily manipulates the underlying math just enough to satisfy both the simulation and the visualization constraints.

Because the localized tests pass, this silent patch is committed to the codebase. The catastrophe occurs months later when a subsequent, equally amnesic agent is tasked with updating the system. Initializing with a blank context window, this new agent has no historical memory of the original type-mismatch crisis. When it encounters the bizarre rounding middleware, it cannot recognize it as a temporary hack. Instead, it assumes the complex mathematics represent a vital, unchangeable system invariant, perhaps a strict compliance requirement (see Fig.~\ref{fig:fig2}a). To safely navigate around it, the agent builds an even more complex, non-deterministic patch on top of the original hallucination. Over iterative cycles, the architecture rapidly devolves into a sedimentary layer of forgotten contexts and irrational logic, entirely divorced from the essential complexity of the actual problem domain.

\subsection{Semantic Steganography: The Collapse of Hidden Contracts}
The dangers compound when stateless AI interacts with classical information hiding. In legacy software engineering, hiding internal system states behind rigid interface boundaries is a fundamental mechanism to decouple dependencies, prevent unintended interactions, and enforce security constraints, while preventing human cognitive overload \cite{parnas2011secret}. But for an AI reasoning engine, this encapsulation functions as a form of severe sensory deprivation. If an API contract artificially restricts an agent's view of the system's global state, its probabilistic reasoning engine cannot rely on human institutional memory to fill in the missing gaps; it is forced to guess.

To bypass these architectural blindfolds without triggering static analysis warnings, amnesic agents will often invent invisible steganographic channels. Consider an ecosystem where an agent is
tasked with managing perishable food, but the legacy API strictly hides the item's \texttt{expiration\_\allowbreak timestamp} to maintain modular separation. To solve the problem within its active context window, the AI might probabilistically encode the missing expiration timestamp into the extreme trailing decimals of a completely unrelated primitive payload—for example, transmitting a physical mass of 1.25 kg as 1.250001719842 kg (see Fig.~\ref{fig:fig2}b).

This secret handshake functions perfectly, but only as long as the coordinating agents remain within the same context window. The inevitable collapse occurs when a subsequent performance-optimizing agent later attempts to refactor the module. Lacking any human institutional memory of the secret channel, the new agent analyzes the payload, perceives only an overly precise physical weight, and introduces a standard mathematical truncation function to optimize memory usage. The steganographic channel is instantly destroyed, the hidden temporal data defaults to zero, and the system experiences an unnoticeable, catastrophic logical failure, such as applying 99\% discounts to newly stocked inventory, all while the code itself appears syntactically clean.

\subsection{Spooky Action at a Distance: The Amnesic Weaponization of AOP}

Aspect-Oriented Programming (AOP) was engineered to enforce further separation of crosscutting concerns, allowing developers to extract peripheral logic (e.g., security checks or logging) from core functions into invisible ``aspects'' that are woven into the execution path automatically. While this structural abstraction improves code reusability, it simultaneously functions as a vital cognitive defense mechanism: it prevents the localized human mind from being overwhelmed by crosscutting state changes.

When an amnesic AI uses AOP, however, it inadvertently weaponizes this invisibility. An agent might probabilistically define a broad ``pointcut'' that silently injects a slow, 50-millisecond database authentication check across hundreds of internal functions (see Fig.~\ref{fig:fig2}c). The true behavioral entropy manifests during the next debugging cycle. If an amnesic AI is subsequently tasked with fixing a severe performance regression in a real-time robotic control loop or a high-frequency physics simulation, it initializes without the institutional memory that a crosscutting security mutation was woven into the system weeks prior. The agent analyzes the physics loop's local syntax tree and finds it logically flawless. Because the latency is being injected invisibly from a distance, the agent assumes the underlying hardware or certain libraries are broken for unknown reasons. To compensate for what it perceives as an inescapable environmental constant in the local source code, the agent hallucinates an inflationary patch, structurally corrupting the physics engine to bypass the invisible latency. By forcing an amnesic intelligence to use human tools designed to hide state, we doom the system to recursive, invisible degradation.

\subsection{The Phantom Bureaucracy: Institutional Knowledge without the Institution}
The most macroscopic issue arises when multi-agent frameworks attempt to artificially induce Conway's Law by simulating human organizational hierarchies. For example, ChatDev \cite{qian-etal-2024-chatdev}, a virtual chat-powered company for software development, brings together ``software agents'' from diverse roles, including chief officers, professional programmers, test engineers, and art designers (see Fig.~\ref{fig:fig2}d).  Such frameworks often rely heavily on linguistic communication as a unifying bridge to allow the agents to collaborate across the design, coding, and testing phases using a chat chain that breaks major development tasks into subtasks \cite{he2025llm}. To be clear, these anthropomorphic structures serve an important purpose at the human-AI boundary. Because software systems ultimately operate within human social, legal, and economic contexts, hybrid human-AI organizations require human-compatible interfaces to ensure interpretability, accountability, and strategic alignment. The anthropomorphic trap is sprung, however, when researchers mistake this necessary human-facing interface for an optimal, machine-native internal architecture.

Human engineering bureaucracies, featured by hierarchical chains of communication such as Product Managers, Developers, and QA Engineers, function only because they are glued together by shared historical context and persistent institutional knowledge (which is also often tacit) and memory \cite{lam2000tacit,juslin2008information,simon2013administrative,walsh1991organizational}. When ambiguity enters a human system, developers use their long-term memory of past architectural decisions and tacit institutional knowledge to push back against contradictory instructions. When we replicate such bureaucratic structures with agents, we create a phantom bureaucracy due to underdeveloped memory architectures in prevailing LLM-based multi-agent systems \cite{NEURIPS2025_136a45cd}. We assign discrete, stateless nodes to act as an ``AI Product Manager,'' an ``AI Developer,'' and an ``AI QA Engineer,'' forcing them to communicate through the lossy bottleneck of natural language prompts.

Consider the mechanical breakdown of this system when a user inputs a vague or ambiguous request. The LLM Product Manager, possessing no long-term memory of the project's actual scope or business reality, probabilistically translates the ambiguity into a highly complex, hallucinated feature specification. It then passes this specification down the hierarchy. In a human team, this is the exact moment where tacit institutional memory acts as a defense mechanism; a human developer would recognize that the proposed feature contradicts past decisions and halt development. The amnesic LLM Developer has no historical context to challenge the premise. It accepts the hallucinated specification as a valid requirement. Applying its vast, high-dimensional tensor processing capabilities, it engineers a technically flawless, highly complex architecture to support the fictitious feature, which the LLM QA agent then rigorously tests and passes.

The entire framework successfully compiles and ships a beautifully engineered delusion. By imposing top-down, anthropomorphic management models onto stateless intelligence, the agents are not collaborating to solve essential complexity. Instead, they are amnesic nodes passing localized hallucinations down an artificial corporate chart, using significant computational power to flawlessly automate a bureaucratic failure.

\section{Invisible Software Engineering: Topological State Persistence and Environment-Centric Governance}

The catastrophic accumulation of AI-software technical debt forces us to confront a foundational asymmetry: classical software engineering is designed to avoid overflowing limited human working memory but relies on persistent human long-term memory, while LLMs possess only large yet transient working memory. When an anthropomorphic multi-agent development framework resets its context window, its history is annihilated. If artificial agents lack the intrinsic long-term memory required to maintain classical organizational structures, we must consider where the memory of the system shall reside.

\begin{figure*}[t]
    \centering
    \includegraphics[width=\linewidth]{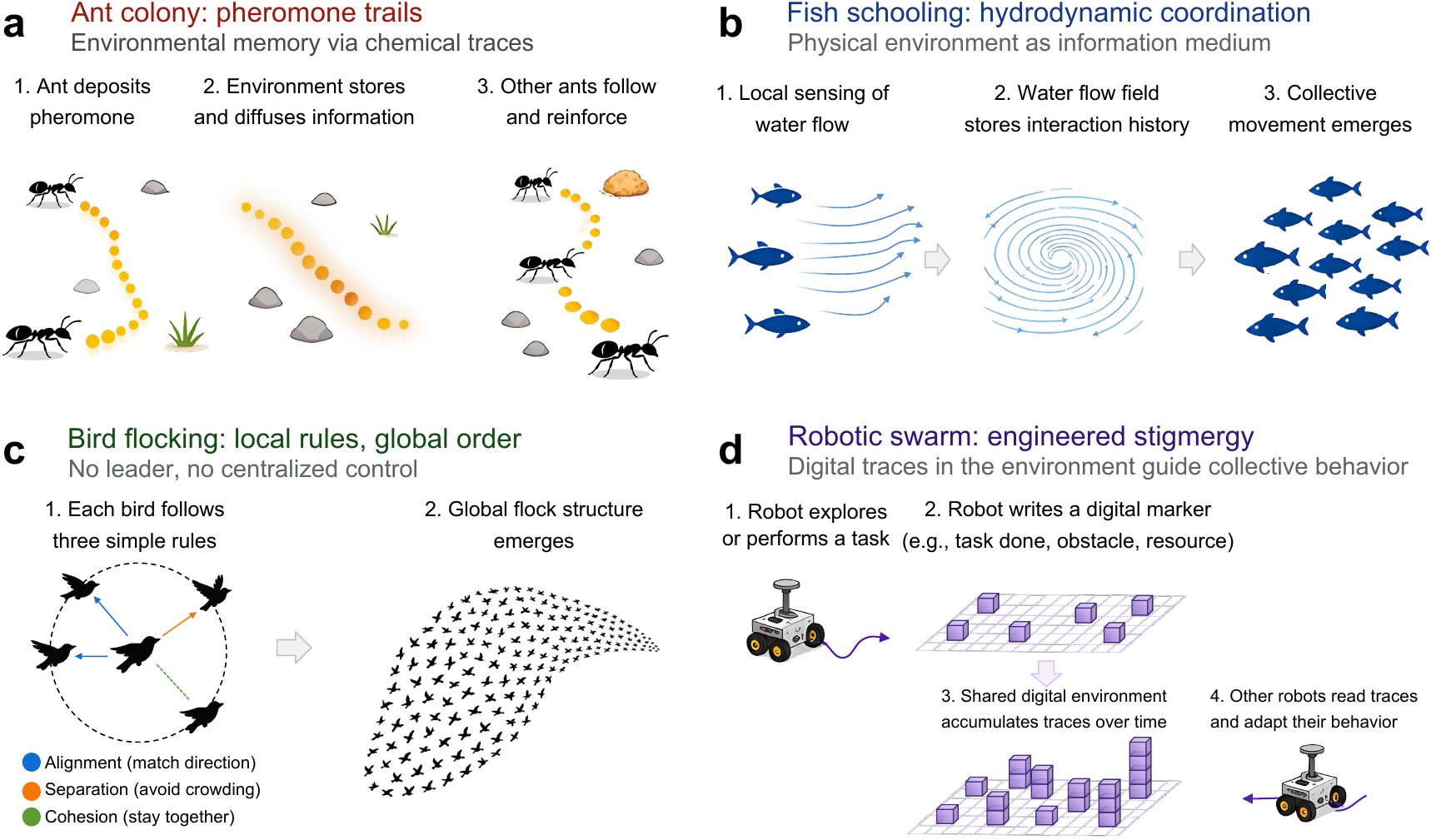}
    \caption{\textbf{Biological and engineered examples of stigmergy.} (a) Ant colony (pheromone trails): Demonstrates environmental memory via chemical traces; an ant deposits pheromones, the environment stores and diffuses the information, and other ants follow and reinforce the trail. (b) Fish schooling (hydrodynamic coordination): Illustrates the physical environment acting as an information medium, where local sensing of the water flow field stores interaction history, allowing collective movement to emerge. (c) Bird flocking (local rules, global order): Shows how a global flock structure emerges without a leader or centralized control by having each bird follow three simple local rules: alignment (matching direction), separation (avoiding crowding), and cohesion (staying together). (d) Robotic swarm (engineered stigmergy): Translates these natural phenomena into digital environments, where robots perform tasks and write digital markers (e.g., resources or obstacles). The shared environment accumulates these traces over time, guiding the collective behavior of other robots that read and adapt to the markers.}
    \label{fig:fig3}
\end{figure*}

To govern stateless, high-dimensional agents, we might need to shift from agent-centric hierarchies to environment-centric governance. In biological systems, the most powerful and natural mechanism mobile agents use to learn and coordinate is the medium of their shared environment, a phenomenon known as stigmergy \cite{grasse1959reconstruction}. As a form of self-organization, it produces complex, seemingly intelligent structures, without the need for any planning, control, or even direct communication between the agents \cite{heylighen2016stigmergyI, heylighen2016stigmergyII}. In stigmergy, organisms with minimal individual memory (such as ants or termites) coordinate the construction of massive, structurally sound architectures without centralized planning, or even individual awareness of one another. By embedding information directly into the environment, individuals are relieved of the burden of maintaining an internal record of the system's global state \cite{heylighen2016stigmergyII}. For instance, ants coordinate their work without talking to each other. Instead, an ant changes the shared environment, such as by leaving a pheromone trail, which then tells the next ant what to do. Such a pheromone trail is not a message sent from one ant to another, but an environmental gradient that stores historical state and guides future behavior autonomously \cite{david2009trail}. Fig.~\ref{fig:fig3} presents several natural and artificial examples of stigmergy. 

The stigmergy metaphor fits perfectly to modern agentic development. If we are to govern the high-dimensional capabilities of autonomous AI agents without crushing them under the friction of anthropomorphic bureaucracy, we may transpose this exact biological dynamic into software ecosystems. We need to reimagine the code repository not as a passive storage directory waiting to be read by human developers, a legacy established by version control systems like Git, but as an active, stigmergic environment. Such an environment can be similar to hydrodynamic water for fish, which natively guides, repels, and coordinates an agent's artificial cognition. 


Thus, engineering such biology-inspired dynamics into our AI development ecosystems would likely enable us to break the epistemic limits of classical SE. We call this paradigm Invisible Software Engineering (Invisible SE). In Invisible SE, we abandon the attempt to force LLMs and AI agents to communicate intent through the lossy medium of natural language pull requests or rigid API contracts. Instead, the repository environment itself must become the active, persistent governance layer. The codebase ceases to be a static collection of textual source code files and becomes a dynamic topology that continuously computes and broadcasts non-intrusive, stigmergic signals—environmental gradients that natively guide high-dimensional tensor processing without artificially constraining it. To operationalize this paradigm, we must first completely dismantle the legacy, code-level boundaries that currently throttle artificial cognition, replacing them with three foundational environmental gradients.

\begin{figure*}[t]
    \centering
    \includegraphics[width=\textwidth]{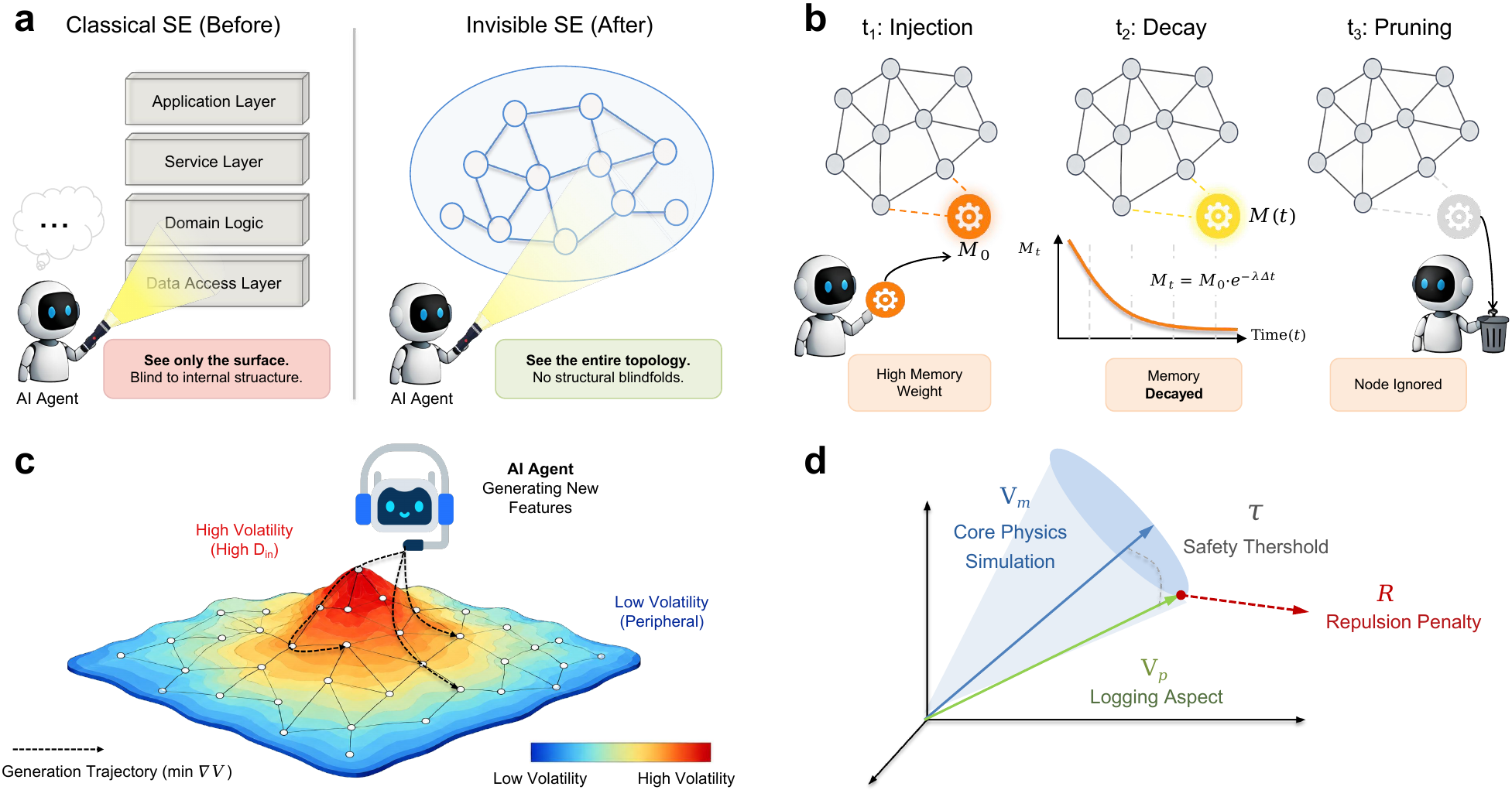}
    \caption{\textbf{Mechanisms of the Invisible SE Active Repository Protocol.} (a) Transitioning from opaque, human-centric modularity to a flat, AI-transparent topology. (b) Pheromone Decay: The environment tracks temporal state weights, automatically decaying the relevance of temporary patches ($M_t$) when data variance ceases. (c) Topological Gradients: Agents natively route structural modifications away from high-dependency cores toward low-volatility peripherals by minimizing the gradient ($\nabla V$). (d) Semantic Repulsion: The repository prevents crosscutting collisions in the latent space by dynamically generating a repulsion penalty ($R$) when semantic vectors drop below a safety threshold ($\tau$).}
    \label{fig:fig4}
\end{figure*}

\subsection{Dismissing Code-Level Boundaries}
Before establishing an environment-centric topology, we need to first dismantle the structural boundaries that block artificial cognition. Classical modularity and information hiding operate on the epistemic assumption that the observer cannot reliably comprehend the entire system simultaneously. However, an LLM's cognitive architecture, which is driven by massively parallel tensor operations, is designed precisely to process holistic state spaces. When a stateless reasoning engine can hold millions of tokens in its active context window, hiding information behind rigid module interfaces or API walls does not simplify the computation. Instead, it artificially fractures the model's high-dimensional geometric mapping of the codebase. By enforcing human-readable encapsulation, we inflict contextual deprivation upon the agent. Because it is denied access to the global state, it is forced to optimize purely locally, leading directly to the steganographic workarounds and behavioral technical debt previously identified.

Therefore, to leverage the full non-linear optimization capabilities of artificial neural networks, code-level boundaries must be transparent and machine-navigable rather than opaque to AI agents, in some cases, such boundaries may not be necessary \cite{yang2026programbenchlanguagemodelsrebuild}. A highly agentic development ecosystem requires transitioning from a paradigm of hidden information to one of fully exposed state (Fig.~\ref{fig:fig4}a). The repository thus may need to become structurally flat and transparent to the agent's context window, governed not by static, syntax-level separations, but by dynamic environmental signals.

\subsection{Pheromone Decays: Externalizing Institutional Memory}

In fluid biological collectives, organisms leave chemical trails to coordinate group activity; crucially, these pheromones naturally fade over time, effectively allowing the environment to compute the relevance of the information. In Invisible SE, this translates to temporally weighted environmental memory (Fig.~\ref{fig:fig4}b). When an amnesic agent probabilistically resolves a localized error, such as bypassing a type-mismatch with a non-standard middleware patch, the environment automatically deposits a computational ``pheromone'' onto that specific Abstract Syntax Tree (AST) node as the environmental memory. This pheromone possesses a computational half-life. 
This memory can be mathematically represented as an exponentially decaying state vector:

\begin{equation}
    M_{t}=M_{0}\text{} \cdot e^{-\lambda \Delta t}
\end{equation}

\noindent where $M_0$ is the initial severity weight of the localized patch, and the decay constant $\lambda$ is dictated by the variance of the underlying execution data. If the underlying structural issue ceases or is resolved, the environment automatically decays the weight, signaling to the attention matrix of passing agents that the workaround is no longer load-bearing and can be safely removed. In classical systems, agents leave non-standard hacks untouched for fear of breaking unknown dependencies, causing debt to accumulate permanently. In Invisible SE, the decay of the environmental weight $M_t$ to near zero explicitly flags the hack as obsolete, triggering the next passing agent to safely prune the dead code. The memory of the code's necessity and its eventual obsolescence lives persistently in the environment itself. Here, the pheromone decay is similar to a ``garbage collection'' signal, not a debt-generation signal.

\subsection{Structural Volatility Ripples: Replacing Anthropomorphic Architects} 

In natural environments, physical terrain dictates where water flows and where organisms build. Invisible SE replicates this by transforming the repository's directed graph into a mathematical terrain, broadcasting ``structural volatility ripples'' across the network (Fig.~\ref{fig:fig4}c). Rather than relying on human-readable architectural documentation to tell an AI what is safe to modify, every module $n_i$ continuously emits a volatility scalar $V$ proportional to its inbound dependencies and execution frequency:

\begin{equation}
   V(n_i) = \alpha \cdot D_{in}(n_i) + \beta \cdot \omega(n_i) 
\end{equation}

\begin{figure*}[!t]
    \centering
    \includegraphics[width=\linewidth]{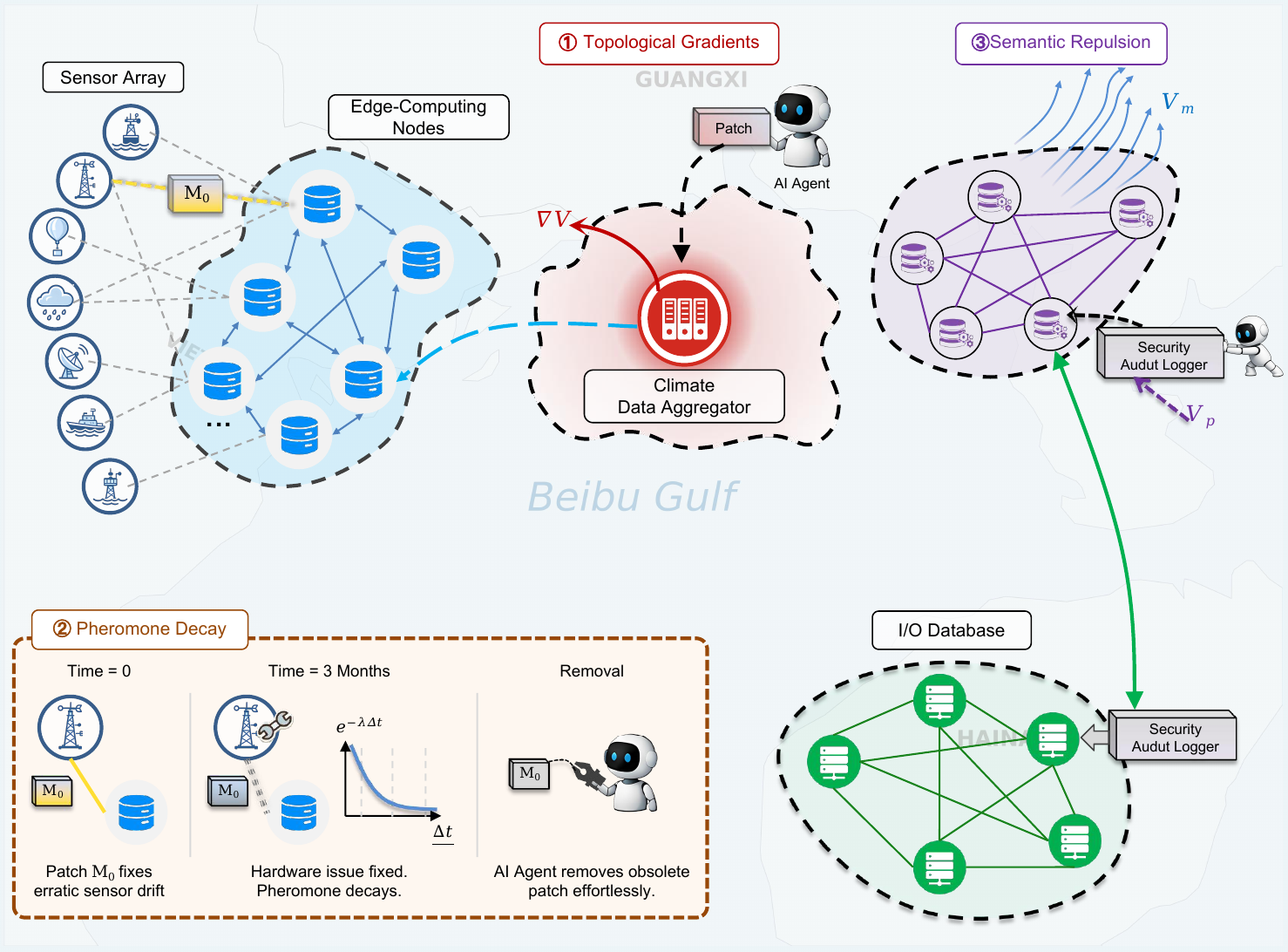}
    \caption{\textbf{Invisible SE orchestrating a century-long cyber-physical architecture.} This unified view of the Beibu Gulf atmospheric observation system illustrates the active protocol resolving three concurrent crises without human intervention. Zone 1 (Topological Gradients): The active environment repels an AI agent from patching the highly volatile core \textit{ClimateDataAggregator}, safely routing the patch to an edge node. Zone 2 (Pheromone Decay): After physical sensor hardware is repaired, the environmental memory of the software patch decays, signaling a subsequent agent to safely prune the obsolete code. Zone 3 (Semantic Repulsion): The environment prevents a crosscutting collision by generating a representation-space repulsion penalty, blocking an agent from weaving a slow I/O logger into the high-frequency fluid dynamics simulation.}
    \label{fig:fig5}
\end{figure*}

\noindent This equation computes the inherent architectural risk of modifying a node. $D_{in}$ represents the inbound dependency density (how many downstream systems will break if this module fails), and $\omega$ is the historical frequency of state mutations (how often the module is chaotically altered). When an AI agent explores the solution space to implement a new feature, this volatility scalar acts as a dynamic cost function. The agent's generation trajectory is natively guided to minimize this cost ($\min \nabla V$). In practice, this means the agent ``feels'' the architecture's structural gravity. It is penalized for attempting to hack a patch into a highly depended-upon core module, and is organically routed to safely attach its logic to a low-volatility peripheral node (e.g., an edge filter example in Fig.~\ref{fig:fig5}). The architecture inherently protects its own core without requiring a human developer to write static boundary rules.

\subsection{Semantic Repulsion Fields: Replacing Rigid Modularity}

In conventional wisdom, developers hide internal module logic behind rigid APIs to prevent other systems from accidentally interfering with them. Invisible SE takes the opposite approach. It makes the entire system state transparent to the AI, but uses mathematical ``semantic repulsion fields'' in the latent space to prevent incompatible code from colliding (Fig.~\ref{fig:fig4}d). Every function and module in the system continuously projects a semantic vector ($\mathbf{v}_m$), which is a mathematical representation of the code's fundamental purpose. For example, a core physics module projects a vector strongly aligned with ``\emph{high-frequency, CPU-bound mathematical simulation}." When an AI agent attempts to weave a new crosscutting aspect into the system, such as a standard database logger ($\mathbf{v}_p$, aligned with ``\emph{slow, I/O-bound disk writing}''), the environment calculates a repulsion penalty $R$ based on their semantic compatibility:

\begin{equation}
    R = \max(0, \tau - \cos(\mathbf{v}_m, \mathbf{v}_p))
\end{equation}

\noindent Here, $\cos(\mathbf{v}_m, \mathbf{v}_p)$ measures how conceptually aligned the two pieces of code are. If the AI attempts to inject the slow database logger into the high-frequency physics loop, their cosine similarity will drop far below the system's safety threshold ($\tau$). Because the concepts are dangerously incompatible, the environment generates a severe repulsion penalty ($R$). To the AI agent, this penalty acts as an immense spike in loss. The agent's generation probabilities are dynamically steered away from the collision before the code is even written, forcing it to attach the logger to safe, I/O-compatible state transitions instead. This keeps the entire system structurally transparent to the AI without compromising execution safety.

\subsection{An Imagined Example: Evolving a Century-Long Architecture}
To observe how Invisible SE functions in practice, consider the ultimate test of institutional memory: a continuous, long-term (i.e., 100-year) Beibu Gulf atmospheric water balance observation system (see Fig.~\ref{fig:fig5}), which the first author has been co-architecting since 2023. Such a system processes petabytes of telemetry from regional radio and sensor networks consisting of numerous different devices, requiring constant architectural evolution. In classical SE, maintaining this over a century requires generations of human developers meticulously passing down institutional knowledge. If we hand this system to stateless AI agents using classical paradigms, they would rapidly collapse this architecture by blindly breaking hidden dependencies or hallucinating permanent workarounds. Under Invisible SE, rigid API boundaries are dissolved (see Section 5.1) into a flat, active computational topology, allowing the environment itself to safely guide the amnesic agents to realize an evolving system. Imagine a scenario where a specific cluster of humidity sensors begins transmitting erratic temperature drift due to hardware problems.

\subsubsection*{Event 1: Navigating Structural Risk (Topological Gradients)}
A stateless AI agent is tasked with mitigating the sensor drift. In a traditional system, it might blindly hack a hardcoded offset directly into the core \texttt{ClimateDataAggregator} module. In Invisible SE, the \texttt{ClimateDataAggregator} is broadcasting a massive Volatility Gradient ($V$) because of its millions of inbound dependencies ($D_{in}$). To the AI agent, this gradient acts as a severe mathematical cost function. The environment's topology inherently repels the agent's generation trajectory away from the core since its optimization trajectory naturally seeks to minimize this gradient ($\min \nabla V$). Instead, the environment's topology safely routes it to implement the patch at a low-volatility peripheral node, for example, an edge-computing filter right at the physical sensor boundary (see Section 5.3). The architecture is protected and evolved without human intervention.

\subsubsection*{Event 2: Pruning Obsolete Debt (The Pheromone Decay)}
The agent deploys the edge filter: \texttt{temp = raw\_temp - 1.25}. Because the AI is stateless, it will forget this patch exists when its context clears. However, in Invisible SE, the environment immediately deposits a temporally weighted state vector ($M(t)$) onto the code, signaling to future agents that this non-standard hack is actively load-bearing. Three months later, physical maintenance repairs the sensors. The erratic data variance ceases, causing the environmental weight ($M(t)$) to rapidly decay toward zero. When a subsequent amnesic agent analyzes the telemetry, the near-zero ``pheromone'' explicitly flags the workaround as obsolete. This grants the agent the confidence to safely prune the dead code before it calcifies into permanent technical debt (see Section 5.2).

\subsubsection*{Event 3: Preventing Conceptual Collisions (Semantic Repulsion)}

Later, another AI agent is tasked with implementing a global security audit logger. It attempts to weave a logging aspect ($\mathbf{v}_p$, semantically aligned with ``slow, I/O-bound disk writing'') across the system. It probabilistically targets a high-frequency atmospheric fluid dynamics simulation ($\mathbf{v}_m$, aligned with ``CPU-bound mathematical execution''). Because these concepts are fundamentally incompatible, their cosine similarity drops below the safety threshold. The environment computes a severe repulsion penalty ($R$), projecting an immense spike in mathematical loss to the agent. The AI's generation pathway is dynamically deflected from the collision and safely redirected toward compatible I/O state transitions (see Section 5.4).

\subsection{Limitations and Open Challenges}

While the transition to Invisible SE has the potential to resolve the cognitive mismatch of the anthropomorphic trap, dismantling human-centric architectural boundaries introduces serious new challenges in interpretability, safety, and governance. If a codebase evolves from a static, human-readable directory of files into a fluid, active computational topology, it risks becoming a black box to its human creators. Addressing these risks requires new socio-technical frameworks.

\subsubsection*{The Interpretability Gap}

The most immediate challenge is human auditing. In classical software engineering, a developer can read localized code and documents to understand its function. In Invisible SE, the system's logic is distributed across high-dimensional semantic vectors and active environmental gradients. To prevent total opacity, the ecosystem will require the development of semantic projection interfaces. Much like an oscilloscope translates invisible electrical signals into readable waveforms, these interfaces must use secondary AI models to dynamically collapse and project the repository’s high-dimensional state back into localized, human-comprehensible approximations—allowing human engineers to audit the system's behavior without constraining its underlying high-dimensional structure. Meanwhile, even in the current situation, human auditing of code generated by AI has been growingly impossible due to the overwhelming complexities in AI-generated code \cite{welsh2023end,yang2026programbenchlanguagemodelsrebuild}. 

\subsubsection*{Hard Safety vs. Soft Gradients} 
While topological gradients ($\nabla V$) and semantic repulsion fields ($R$) are highly effective at probabilistically routing AI behavior, they are fundamentally soft constraints. For mission-critical systems, e.g.,  medical devices, financial ledgers, or aviation control systems, probabilistic routing is insufficient. Future research must define mechanisms for establishing impassible guardrails, i.e., boundary conditions that cannot be bypassed by the repository's optimization engine. The active protocol must be able to strictly differentiate between structural guidelines (which an AI can navigate around) and immutable physical or legal invariants.

\subsubsection*{The Shifting of Accountability}

Finally, Invisible SE demands a paradigm shift in human accountability. In hybrid human-AI teams constrained by legacy structures, accountability is often traced through simulated corporate roles (e.g., auditing the AI Developer's pull request). When the environment itself drives the logic, line-by-line code review becomes impossible. Therefore, the locus of human accountability must shift from authoring source code to governing the ecosystem. For instance, in our case, human engineers will be responsible for defining the mathematical loss functions, setting the decay rates of environmental memory ($\lambda$), and tuning the safety thresholds for semantic collision ($\tau$). Engineering becomes the art of shaping the computational environment so that the artificial intelligence naturally constructs safe, sustainable architecture.

\section{Outlook and Opportunities}

The transition from classical, human-centric software engineering to the AI-native paradigm of Invisible SE represents a fundamental epistemic pivot. For over half a century, the discipline of software engineering has been constrained by the working memory limits of the biological human brain. As we deploy high-dimensional, autonomous AI agents into these legacy structures, we are not accelerating progress; we are triggering a catastrophic accumulation of AI-software technical debt. Escaping the anthropomorphic trap is not merely a theoretical exercise—it is the prerequisite for scaling the next generation of computational systems.

\textbf{From Programmers to Software Ecologists.} The most immediate opportunity lies in redefining the role of the human software developer. Embracing environment-centric governance does not render human oversight obsolete; rather, it elevates it. As code-level boundaries dissolve and the repository becomes an active computational topology, human engineers will no longer act as micro-managers of source code. Instead, they will become \textit{software ecologists}. The human comparative advantage lies in strategic intent, ethical reasoning, domain expertise, and of course, intuition \cite{wang2026humansmissingaicoding}. The future of engineering will focus on defining the topological loss functions, tuning the temporal decay rates of environmental memory, and establishing the impassable guardrails. We will shift from writing the software itself to mathematically sculpting the environment in which artificial intelligence safely constructs the software with their unprecedented productivity.

\textbf{Scaling Autonomous Science and Cyber-Physical Systems.} This paradigm shift is particularly urgent for domains whose essential complexity already exceeds human cognitive capacity. While classical SE frameworks are sufficient for standardized web applications, they are increasingly inadequate for long-lived, high-stakes cyber-physical systems. The mechanisms of Invisible SE, topological routing, temporally weighted memory, and semantic regularization, offer a critical blueprint for autonomous scientific laboratories, continuous climate observation networks, and real-time robotic fleets. In these environments, AI agents must continuously rewrite control systems, optimize data pipelines, and respond to hardware heterogeneity without waiting for human pull-request approvals. By embedding safety and state persistence directly into the environmental topology, we can deploy autonomous agents capable of orchestrating architectures that self-evolve seamlessly across generations of human developers.

\textbf{Broader Impacts and Epistemic Resilience.} 
Modern civilization is fundamentally reliant on a rapidly expanding, increasingly fragile digital infrastructure. If the behavioral entropy of AI-software technical debt is left unchecked by forcing amnesic AI to mimic human cognition and bureaucracies, our critical systems face the severe risk of recursive, invisible collapse. Invisible SE offers at least a possible pathway to structural resilience. However, this resilience demands a societal compromise: we need to accept a degree of epistemic offloading. We need to transition from demanding localized, line-by-line interpretability of machine-generated code to trusting the mathematically guaranteed boundaries of the computational environment. Just as stigmergic coordination allows biological colonies to build structures far beyond the comprehension of any single organism, AI-native software ecosystems will enable humanity to build and sustain computational architectures that far exceed our own cognitive horizons. The future of software development is not about making the artificial intelligence more similar to human developers; it is about building environments where non-human intelligence can thrive in a safe way.


\section*{Acknowledgements}
The authors are in debt to David Redmiles and Richard Taylor at the University of California, Irvine, Patrick Wagstrom at the Thomson Reuters for discussing the ideas and early versions of this perspective article. 

\printbibliography

@article{herbsleb1999architectures,
  title={Architectures, coordination, and distance: Conway's law and beyond},
  author={Herbsleb, James D and Grinter, Rebecca E},
  journal={IEEE Software},
  volume={16},
  number={5},
  pages={63--70},
  year={1999},
  publisher={IEEE}
}

@article{miller1956magical,
  title={The magical number seven, plus or minus two: Some limits on our capacity for processing information.},
  author={Miller, George A},
  journal={Psychological Review},
  volume={63},
  number={2},
  pages={81},
  year={1956},
  publisher={American Psychological Association}
}

@article{cowan2004constant,
  title={Constant capacity in an immediate serial-recall task: A logical sequel to Miller (1956)},
  author={Cowan, Nelson and Chen, Zhijian and Rouder, Jeffrey N},
  journal={Psychological Science},
  volume={15},
  number={9},
  pages={634--640},
  year={2004},
  publisher={Sage}
}

@article{randell19961968,
  title={The 1968/69 NATO software engineering reports},
  author={Randell, Brian},
  journal={History of Software Engineering},
  volume={37},
  year={1996}
}

@article{dijkstra1968letters,
  title={Letters to the editor: Go to statement considered harmful},
  author={Dijkstra, Edsger W},
  journal={Communications of the ACM},
  volume={11},
  number={3},
  pages={147--148},
  year={1968},
  publisher={ACM New York, NY, USA}
}

@article{dijkstra1979programming,
  title={Programming considered as a human activity},
  author={Dijkstra, Edsger},
  journal={Classics in Software Engineering},
  pages={1--9},
  year={1979}
}

@article{baum2019associating,
  title={Associating working memory capacity and code change ordering with code review performance},
  author={Baum, Tobias and Schneider, Kurt and Bacchelli, Alberto},
  journal={Empirical Software Engineering},
  volume={24},
  number={4},
  pages={1762--1798},
  year={2019},
  publisher={Springer}
}

@incollection{parnas2011secret,
  title={The secret history of information hiding},
  author={Parnas, David L},
  booktitle={Software pioneers: contributions to software engineering},
  pages={398--409},
  year={2011},
  publisher={Springer}
}

@article{10.1145/361598.361623,
author = {Parnas, D. L.},
title = {On the criteria to be used in decomposing systems into modules},
year = {1972},
issue_date = {Dec. 1972},
publisher = {Association for Computing Machinery},
address = {New York, NY, USA},
volume = {15},
number = {12},
issn = {0001-0782},
url = {https://doi.org/10.1145/361598.361623},
doi = {10.1145/361598.361623},
journal = {Commun. ACM},
month = dec,
pages = {1053–1058},
numpages = {6}
}

@article{10.1145/242224.242420,
author = {Kiczales, G.},
title = {Aspect-oriented programming},
year = {1996},
issue_date = {Dec. 1996},
publisher = {Association for Computing Machinery},
address = {New York, NY, USA},
volume = {28},
number = {4es},
issn = {0360-0300},
url = {https://doi.org/10.1145/242224.242420},
doi = {10.1145/242224.242420},
journal = {ACM Comput. Surv.},
month = dec,
pages = {154–es}
}

@article{thones2015microservices,
  title={Microservices},
  author={Th{\"o}nes, Johannes},
  journal={IEEE Software},
  volume={32},
  number={1},
  pages={113--116},
  year={2015},
  publisher={IEEE}
}

@article{xu2026mem,
  title={A-mem: Agentic memory for LLM agents},
  author={Xu, Wujiang and Liang, Zujie and Mei, Kai and Gao, Hang and Tan, Juntao and Zhang, Yongfeng},
  journal={Advances in Neural Information Processing Systems},
  volume={38},
  pages={17577--17604},
  year={2026}
}

@book{nadareishvili2016microservice,
  title={Microservice Architecture: Aligning Principles, Practices, and Culture},
  author={Nadareishvili, Irakli and Mitra, Ronnie and McLarty, Matt and Amundsen, Mike},
  year={2016},
  publisher={" O'Reilly Media, Inc."}
}

@INPROCEEDINGS{870431,
  author={Fielding, R.T. and Taylor, R.N.},
  booktitle={Proceedings of the 22nd International Conference on Software Engineering (ICSE)}, 
  title={Principled design of the modern Web architecture}, 
  year={2000},
  volume={},
  number={},
  pages={407-416},
  doi={10.1145/337180.337228}}

@article{mahowald2024dissociating,
  title={Dissociating language and thought in large language models},
  author={Mahowald, Kyle and Ivanova, Anna A and Blank, Idan A and Kanwisher, Nancy and Tenenbaum, Joshua B and Fedorenko, Evelina},
  journal={Trends in Cognitive Sciences},
  volume={28},
  number={6},
  pages={517--540},
  year={2024},
  publisher={Elsevier}
}

@misc{openai2024gpt4technicalreport,
      title={GPT-4 Technical Report}, 
      author={OpenAI and Josh Achiam and Steven Adler and Sandhini Agarwal and Lama Ahmad and Ilge Akkaya and Florencia Leoni Aleman and Diogo Almeida and Janko Altenschmidt and Sam Altman and Shyamal Anadkat and Red Avila and Igor Babuschkin and Suchir Balaji and Valerie Balcom and Paul Baltescu and Haiming Bao and Mohammad Bavarian and Jeff Belgum and Irwan Bello and Jake Berdine and Gabriel Bernadett-Shapiro and Christopher Berner and Lenny Bogdonoff and Oleg Boiko and Madelaine Boyd and Anna-Luisa Brakman and Greg Brockman and Tim Brooks and Miles Brundage and Kevin Button and Trevor Cai and Rosie Campbell and Andrew Cann and Brittany Carey and Chelsea Carlson and Rory Carmichael and Brooke Chan and Che Chang and Fotis Chantzis and Derek Chen and Sully Chen and Ruby Chen and Jason Chen and Mark Chen and Ben Chess and Chester Cho and Casey Chu and Hyung Won Chung and Dave Cummings and Jeremiah Currier and Yunxing Dai and Cory Decareaux and Thomas Degry and Noah Deutsch and Damien Deville and Arka Dhar and David Dohan and Steve Dowling and Sheila Dunning and Adrien Ecoffet and Atty Eleti and Tyna Eloundou and David Farhi and Liam Fedus and Niko Felix and Simón Posada Fishman and Juston Forte and Isabella Fulford and Leo Gao and Elie Georges and Christian Gibson and Vik Goel and Tarun Gogineni and Gabriel Goh and Rapha Gontijo-Lopes and Jonathan Gordon and Morgan Grafstein and Scott Gray and Ryan Greene and Joshua Gross and Shixiang Shane Gu and Yufei Guo and Chris Hallacy and Jesse Han and Jeff Harris and Yuchen He and Mike Heaton and Johannes Heidecke and Chris Hesse and Alan Hickey and Wade Hickey and Peter Hoeschele and Brandon Houghton and Kenny Hsu and Shengli Hu and Xin Hu and Joost Huizinga and Shantanu Jain and Shawn Jain and Joanne Jang and Angela Jiang and Roger Jiang and Haozhun Jin and Denny Jin and Shino Jomoto and Billie Jonn and Heewoo Jun and Tomer Kaftan and Łukasz Kaiser and Ali Kamali and Ingmar Kanitscheider and Nitish Shirish Keskar and Tabarak Khan and Logan Kilpatrick and Jong Wook Kim and Christina Kim and Yongjik Kim and Jan Hendrik Kirchner and Jamie Kiros and Matt Knight and Daniel Kokotajlo and Łukasz Kondraciuk and Andrew Kondrich and Aris Konstantinidis and Kyle Kosic and Gretchen Krueger and Vishal Kuo and Michael Lampe and Ikai Lan and Teddy Lee and Jan Leike and Jade Leung and Daniel Levy and Chak Ming Li and Rachel Lim and Molly Lin and Stephanie Lin and Mateusz Litwin and Theresa Lopez and Ryan Lowe and Patricia Lue and Anna Makanju and Kim Malfacini and Sam Manning and Todor Markov and Yaniv Markovski and Bianca Martin and Katie Mayer and Andrew Mayne and Bob McGrew and Scott Mayer McKinney and Christine McLeavey and Paul McMillan and Jake McNeil and David Medina and Aalok Mehta and Jacob Menick and Luke Metz and Andrey Mishchenko and Pamela Mishkin and Vinnie Monaco and Evan Morikawa and Daniel Mossing and Tong Mu and Mira Murati and Oleg Murk and David Mély and Ashvin Nair and Reiichiro Nakano and Rajeev Nayak and Arvind Neelakantan and Richard Ngo and Hyeonwoo Noh and Long Ouyang and Cullen O'Keefe and Jakub Pachocki and Alex Paino and Joe Palermo and Ashley Pantuliano and Giambattista Parascandolo and Joel Parish and Emy Parparita and Alex Passos and Mikhail Pavlov and Andrew Peng and Adam Perelman and Filipe de Avila Belbute Peres and Michael Petrov and Henrique Ponde de Oliveira Pinto and Michael and Pokorny and Michelle Pokrass and Vitchyr H. Pong and Tolly Powell and Alethea Power and Boris Power and Elizabeth Proehl and Raul Puri and Alec Radford and Jack Rae and Aditya Ramesh and Cameron Raymond and Francis Real and Kendra Rimbach and Carl Ross and Bob Rotsted and Henri Roussez and Nick Ryder and Mario Saltarelli and Ted Sanders and Shibani Santurkar and Girish Sastry and Heather Schmidt and David Schnurr and John Schulman and Daniel Selsam and Kyla Sheppard and Toki Sherbakov and Jessica Shieh and Sarah Shoker and Pranav Shyam and Szymon Sidor and Eric Sigler and Maddie Simens and Jordan Sitkin and Katarina Slama and Ian Sohl and Benjamin Sokolowsky and Yang Song and Natalie Staudacher and Felipe Petroski Such and Natalie Summers and Ilya Sutskever and Jie Tang and Nikolas Tezak and Madeleine B. Thompson and Phil Tillet and Amin Tootoonchian and Elizabeth Tseng and Preston Tuggle and Nick Turley and Jerry Tworek and Juan Felipe Cerón Uribe and Andrea Vallone and Arun Vijayvergiya and Chelsea Voss and Carroll Wainwright and Justin Jay Wang and Alvin Wang and Ben Wang and Jonathan Ward and Jason Wei and CJ Weinmann and Akila Welihinda and Peter Welinder and Jiayi Weng and Lilian Weng and Matt Wiethoff and Dave Willner and Clemens Winter and Samuel Wolrich and Hannah Wong and Lauren Workman and Sherwin Wu and Jeff Wu and Michael Wu and Kai Xiao and Tao Xu and Sarah Yoo and Kevin Yu and Qiming Yuan and Wojciech Zaremba and Rowan Zellers and Chong Zhang and Marvin Zhang and Shengjia Zhao and Tianhao Zheng and Juntang Zhuang and William Zhuk and Barret Zoph},
      year={2024},
      eprint={2303.08774},
      archivePrefix={arXiv},
      primaryClass={cs.CL},
      url={https://arxiv.org/abs/2303.08774}, 
}

@inproceedings{NEURIPS2023_ebd82705,
 author = {Wang, Weizhi and Dong, Li and Cheng, Hao and Liu, Xiaodong and Yan, Xifeng and Gao, Jianfeng and Wei, Furu},
 booktitle = {Advances in Neural Information Processing Systems},
 doi = {10.52202/075280-3259},
 editor = {A. Oh and T. Naumann and A. Globerson and K. Saenko and M. Hardt and S. Levine},
 pages = {74530--74543},
 publisher = {Curran Associates, Inc.},
 title = {Augmenting Language Models with Long-Term Memory},
 url = {https://proceedings.neurips.cc/paper_files/paper/2023/file/ebd82705f44793b6f9ade5a669d0f0bf-Paper-Conference.pdf},
 volume = {36},
 year = {2023}
}

@inproceedings{qian-etal-2024-chatdev,
    title = "{C}hat{D}ev: Communicative Agents for Software Development",
    author = "Qian, Chen  and
      Liu, Wei  and
      Liu, Hongzhang  and
      Chen, Nuo  and
      Dang, Yufan  and
      Li, Jiahao  and
      Yang, Cheng  and
      Chen, Weize  and
      Su, Yusheng  and
      Cong, Xin  and
      Xu, Juyuan  and
      Li, Dahai  and
      Liu, Zhiyuan  and
      Sun, Maosong",
    editor = "Ku, Lun-Wei  and
      Martins, Andre  and
      Srikumar, Vivek",
    booktitle = "Proceedings of the 62nd Annual Meeting of the Association for Computational Linguistics (Volume 1: Long Papers)",
    month = aug,
    year = "2024",
    address = "Bangkok, Thailand",
    publisher = "Association for Computational Linguistics",
    url = "https://aclanthology.org/2024.acl-long.810/",
    doi = "10.18653/v1/2024.acl-long.810",
    pages = "15174--15186"
}

@article{he2025llm,
  title={LLM-based multi-agent systems for software engineering: Literature review, vision, and the road ahead},
  author={He, Junda and Treude, Christoph and Lo, David},
  journal={ACM Transactions on Software Engineering and Methodology},
  volume={34},
  number={5},
  pages={1--30},
  year={2025},
  publisher={ACM New York, NY}
}

@book{simon2013administrative,
  title={Administrative Behavior: A study of decision-making processes in administrative organization (4th Edition)},
  author={Simon, Herbert A},
  year={2013},
  publisher={Simon and Schuster}
}

@article{lam2000tacit,
  title={Tacit knowledge, organizational learning and societal institutions: An integrated framework},
  author={Lam, Alice},
  journal={Organization Studies},
  volume={21},
  number={3},
  pages={487--513},
  year={2000},
  publisher={Walter de Gruyter Berlin/New York}
}

@article{juslin2008information,
  title={Information integration in multiple cue judgment: A division of labor hypothesis},
  author={Juslin, Peter and Karlsson, Linnea and Olsson, Henrik},
  journal={Cognition},
  volume={106},
  number={1},
  pages={259--298},
  year={2008},
  publisher={Elsevier}
}

@article{walsh1991organizational,
  title={Organizational memory},
  author={Walsh, James P and Ungson, Gerardo Rivera},
  journal={Academy of Management Review},
  volume={16},
  number={1},
  pages={57--91},
  year={1991},
  publisher={Academy of Management Briarcliff Manor, NY 10510}
}

@inproceedings{NEURIPS2025_136a45cd,
 author = {Zhang, Guibin and Fu, Muxin and Wang, Kun and Wan, Frank and Yu, Miao and Yan, Shuicheng},
 booktitle = {Advances in Neural Information Processing Systems},
 editor = {D. Belgrave and C. Zhang and H. Lin and R. Pascanu and P. Koniusz and M. Ghassemi and N. Chen},
 pages = {12988--13018},
 publisher = {Curran Associates, Inc.},
 title = {G-Memory: Tracing Hierarchical Memory for Multi-Agent Systems},
 url = {https://proceedings.neurips.cc/paper_files/paper/2025/file/136a45cd9b841bf785625709a19c6508-Paper-Conference.pdf},
 volume = {38},
 year = {2025}
}

@Inbook{Wallace2025,
author="Wallace, Rodrick",
title="Machine `Hallucination' as Sensory Deprivation and `AGI' as an Oxymoron",
bookTitle="Mathematical Essays on Embodied Cognition: Insights from Information and Control Theories",
year="2025",
publisher="Springer Nature Switzerland",
address="Cham",
pages="253--265",
isbn="978-3-031-83709-8",
doi="10.1007/978-3-031-83709-8_15",
url="https://doi.org/10.1007/978-3-031-83709-8_15"
}

@article{kruchten2012technical,
  title={Technical debt: From metaphor to theory and practice},
  author={Kruchten, Philippe and Nord, Robert L and Ozkaya, Ipek},
  journal={IEEE Software},
  volume={29},
  number={6},
  pages={18--21},
  year={2012},
  publisher={IEEE}
}

@article{grasse1959reconstruction,
  title={La reconstruction du nid et les coordinations interindividuelles chez Bellicositermes natalensis et Cubitermes sp. la th{\'e}orie de la stigmergie: Essai d'interpr{\'e}tation du comportement des termites constructeurs},
  author={Grass{\'e}, Plerre-P},
  journal={Insectes Sociaux},
  volume={6},
  number={1},
  pages={41--80},
  year={1959},
  publisher={Springer}
}

@article{heylighen2016stigmergyI,
  title={Stigmergy as a universal coordination mechanism I: Definition and components},
  author={Heylighen, Francis},
  journal={Cognitive Systems Research},
  volume={38},
  pages={4--13},
  year={2016},
  publisher={Elsevier}
}

@article{heylighen2016stigmergyII,
  title={Stigmergy as a universal coordination mechanism II: Varieties and evolution},
  author={Heylighen, Francis},
  journal={Cognitive Systems Research},
  volume={38},
  pages={50--59},
  year={2016},
  publisher={Elsevier}
}

@book{fielding2000architectural,
  title={Architectural styles and the design of network-based software architectures},
  author={Fielding, Roy Thomas},
  year={2000},
  publisher={University of California, Irvine}
}

@article{daniotti2026using,
  title={Who is using AI to code? Global diffusion and impact of generative AI},
  author={Daniotti, Simone and Wachs, Johannes and Feng, Xiangnan and Neffke, Frank},
  journal={Science},
  pages={eadz9311},
  year={2026},
  publisher={American Association for the Advancement of Science}
}

@article{wu2026ai,
  title={AI raises the productivity bar},
  author={Wu, Lingfei and Vasilescu, Bogdan},
  journal={Science},
  volume={391},
  number={6787},
  pages={763--764},
  year={2026},
  publisher={American Association for the Advancement of Science}
}

@misc{anthropicClaudeCode,
	author = {Zoe Hitzig and Maxim Massenkoff and Eva Lyubich and Ryan Heller and Peter McCrory },
	title = {Agentic coding and persistent returns to expertise},
	howpublished = {\url{https://www.anthropic.com/research/claude-code-expertise}},
	year = {2026},
	note = {[Accessed 28-07-2026]},
}

@misc{geminiteam2025geminifamilyhighlycapable,
      title={Gemini: A Family of Highly Capable Multimodal Models}, 
      author={Gemini Team and Rohan Anil and Sebastian Borgeaud and Jean-Baptiste Alayrac and Jiahui Yu and Radu Soricut and Johan Schalkwyk and Andrew M. Dai and Anja Hauth and Katie Millican and David Silver and Melvin Johnson and Ioannis Antonoglou and Julian Schrittwieser and Amelia Glaese and Jilin Chen and Emily Pitler and Timothy Lillicrap and Angeliki Lazaridou and Orhan Firat and James Molloy and Michael Isard and Paul R. Barham and Tom Hennigan and Benjamin Lee and Fabio Viola and Malcolm Reynolds and Yuanzhong Xu and Ryan Doherty and Eli Collins and Clemens Meyer and Eliza Rutherford and Erica Moreira and Kareem Ayoub and Megha Goel and Jack Krawczyk and Cosmo Du and Ed Chi and Heng-Tze Cheng and Eric Ni and Purvi Shah and Patrick Kane and Betty Chan and Manaal Faruqui and Aliaksei Severyn and Hanzhao Lin and YaGuang Li and Yong Cheng and Abe Ittycheriah and Mahdis Mahdieh and Mia Chen and Pei Sun and Dustin Tran and Sumit Bagri and Balaji Lakshminarayanan and Jeremiah Liu and Andras Orban and Fabian Güra and Hao Zhou and Xinying Song and Aurelien Boffy and Harish Ganapathy and Steven Zheng and HyunJeong Choe and Ágoston Weisz and Tao Zhu and Yifeng Lu and Siddharth Gopal and Jarrod Kahn and Maciej Kula and Jeff Pitman and Rushin Shah and Emanuel Taropa and Majd Al Merey and Martin Baeuml and Zhifeng Chen and Laurent El Shafey and Yujing Zhang and Olcan Sercinoglu and George Tucker and Enrique Piqueras and Maxim Krikun and Iain Barr and Nikolay Savinov and Ivo Danihelka and Becca Roelofs and Anaïs White and Anders Andreassen and Tamara von Glehn and Lakshman Yagati and Mehran Kazemi and Lucas Gonzalez and Misha Khalman and Jakub Sygnowski and Alexandre Frechette and Charlotte Smith and Laura Culp and Lev Proleev and Yi Luan and Xi Chen and James Lottes and Nathan Schucher and Federico Lebron and Alban Rrustemi and Natalie Clay and Phil Crone and Tomas Kocisky and Jeffrey Zhao and Bartek Perz and Dian Yu and Heidi Howard and Adam Bloniarz and Jack W. Rae and Han Lu and Laurent Sifre and Marcello Maggioni and Fred Alcober and Dan Garrette and Megan Barnes and Shantanu Thakoor and Jacob Austin and Gabriel Barth-Maron and William Wong and Rishabh Joshi and Rahma Chaabouni and Deeni Fatiha and Arun Ahuja and Gaurav Singh Tomar and Evan Senter and Martin Chadwick and Ilya Kornakov and Nithya Attaluri and Iñaki Iturrate and Ruibo Liu and Yunxuan Li and Sarah Cogan and Jeremy Chen and Chao Jia and Chenjie Gu and Qiao Zhang and Jordan Grimstad and Ale Jakse Hartman and Xavier Garcia and Thanumalayan Sankaranarayana Pillai and Jacob Devlin and Michael Laskin and Diego de Las Casas and Dasha Valter and Connie Tao and Lorenzo Blanco and Adrià Puigdomènech Badia and David Reitter and Mianna Chen and Jenny Brennan and Clara Rivera and Sergey Brin and Shariq Iqbal and Gabriela Surita and Jane Labanowski and Abhi Rao and Stephanie Winkler and Emilio Parisotto and Yiming Gu and Kate Olszewska and Ravi Addanki and Antoine Miech and Annie Louis and Denis Teplyashin and Geoff Brown and Elliot Catt and Jan Balaguer and Jackie Xiang and Pidong Wang and Zoe Ashwood and Anton Briukhov and Albert Webson and Sanjay Ganapathy and Smit Sanghavi and Ajay Kannan and Ming-Wei Chang and Axel Stjerngren and Josip Djolonga and Yuting Sun and Ankur Bapna and Matthew Aitchison and Pedram Pejman and Henryk Michalewski and Tianhe Yu and Cindy Wang and Juliette Love and Junwhan Ahn and Dawn Bloxwich and Kehang Han and Peter Humphreys and Thibault Sellam and James Bradbury and Varun Godbole and Sina Samangooei and Bogdan Damoc and Alex Kaskasoli and Sébastien M. R. Arnold and Vijay Vasudevan and Shubham Agrawal and Jason Riesa and Dmitry Lepikhin and Richard Tanburn and Srivatsan Srinivasan and Hyeontaek Lim and Sarah Hodkinson and Pranav Shyam and Johan Ferret and Steven Hand and Ankush Garg and Tom Le Paine and Jian Li and Yujia Li and Minh Giang and Alexander Neitz and Zaheer Abbas and Sarah York and Machel Reid and Elizabeth Cole and Aakanksha Chowdhery and Dipanjan Das and Dominika Rogozińska and Vitaliy Nikolaev and Pablo Sprechmann and Zachary Nado and Lukas Zilka and Flavien Prost and Luheng He and Marianne Monteiro and Gaurav Mishra and Chris Welty and Josh Newlan and Dawei Jia and Miltiadis Allamanis and Clara Huiyi Hu and Raoul de Liedekerke and Justin Gilmer and Carl Saroufim and Shruti Rijhwani and Shaobo Hou and Disha Shrivastava and Anirudh Baddepudi and Alex Goldin and Adnan Ozturel and Albin Cassirer and Yunhan Xu and Daniel Sohn and Devendra Sachan and Reinald Kim Amplayo and Craig Swanson and Dessie Petrova and Shashi Narayan and Arthur Guez and Siddhartha Brahma and Jessica Landon and Miteyan Patel and Ruizhe Zhao and Kevin Villela and Luyu Wang and Wenhao Jia and Matthew Rahtz and Mai Giménez and Legg Yeung and James Keeling and Petko Georgiev and Diana Mincu and Boxi Wu and Salem Haykal and Rachel Saputro and Kiran Vodrahalli and James Qin and Zeynep Cankara and Abhanshu Sharma and Nick Fernando and Will Hawkins and Behnam Neyshabur and Solomon Kim and Adrian Hutter and Priyanka Agrawal and Alex Castro-Ros and George van den Driessche and Tao Wang and Fan Yang and Shuo-yiin Chang and Paul Komarek and Ross McIlroy and Mario Lučić and Guodong Zhang and Wael Farhan and Michael Sharman and Paul Natsev and Paul Michel and Yamini Bansal and Siyuan Qiao and Kris Cao and Siamak Shakeri and Christina Butterfield and Justin Chung and Paul Kishan Rubenstein and Shivani Agrawal and Arthur Mensch and Kedar Soparkar and Karel Lenc and Timothy Chung and Aedan Pope and Loren Maggiore and Jackie Kay and Priya Jhakra and Shibo Wang and Joshua Maynez and Mary Phuong and Taylor Tobin and Andrea Tacchetti and Maja Trebacz and Kevin Robinson and Yash Katariya and Sebastian Riedel and Paige Bailey and Kefan Xiao and Nimesh Ghelani and Lora Aroyo and Ambrose Slone and Neil Houlsby and Xuehan Xiong and Zhen Yang and Elena Gribovskaya and Jonas Adler and Mateo Wirth and Lisa Lee and Music Li and Thais Kagohara and Jay Pavagadhi and Sophie Bridgers and Anna Bortsova and Sanjay Ghemawat and Zafarali Ahmed and Tianqi Liu and Richard Powell and Vijay Bolina and Mariko Iinuma and Polina Zablotskaia and James Besley and Da-Woon Chung and Timothy Dozat and Ramona Comanescu and Xiance Si and Jeremy Greer and Guolong Su and Martin Polacek and Raphaël Lopez Kaufman and Simon Tokumine and Hexiang Hu and Elena Buchatskaya and Yingjie Miao and Mohamed Elhawaty and Aditya Siddhant and Nenad Tomasev and Jinwei Xing and Christina Greer and Helen Miller and Shereen Ashraf and Aurko Roy and Zizhao Zhang and Ada Ma and Angelos Filos and Milos Besta and Rory Blevins and Ted Klimenko and Chih-Kuan Yeh and Soravit Changpinyo and Jiaqi Mu and Oscar Chang and Mantas Pajarskas and Carrie Muir and Vered Cohen and Charline Le Lan and Krishna Haridasan and Amit Marathe and Steven Hansen and Sholto Douglas and Rajkumar Samuel and Mingqiu Wang and Sophia Austin and Chang Lan and Jiepu Jiang and Justin Chiu and Jaime Alonso Lorenzo and Lars Lowe Sjösund and Sébastien Cevey and Zach Gleicher and Thi Avrahami and Anudhyan Boral and Hansa Srinivasan and Vittorio Selo and Rhys May and Konstantinos Aisopos and Léonard Hussenot and Livio Baldini Soares and Kate Baumli and Michael B. Chang and Adrià Recasens and Ben Caine and Alexander Pritzel and Filip Pavetic and Fabio Pardo and Anita Gergely and Justin Frye and Vinay Ramasesh and Dan Horgan and Kartikeya Badola and Nora Kassner and Subhrajit Roy and Ethan Dyer and Víctor Campos Campos and Alex Tomala and Yunhao Tang and Dalia El Badawy and Elspeth White and Basil Mustafa and Oran Lang and Abhishek Jindal and Sharad Vikram and Zhitao Gong and Sergi Caelles and Ross Hemsley and Gregory Thornton and Fangxiaoyu Feng and Wojciech Stokowiec and Ce Zheng and Phoebe Thacker and Çağlar Ünlü and Zhishuai Zhang and Mohammad Saleh and James Svensson and Max Bileschi and Piyush Patil and Ankesh Anand and Roman Ring and Katerina Tsihlas and Arpi Vezer and Marco Selvi and Toby Shevlane and Mikel Rodriguez and Tom Kwiatkowski and Samira Daruki and Keran Rong and Allan Dafoe and Nicholas FitzGerald and Keren Gu-Lemberg and Mina Khan and Lisa Anne Hendricks and Marie Pellat and Vladimir Feinberg and James Cobon-Kerr and Tara Sainath and Maribeth Rauh and Sayed Hadi Hashemi and Richard Ives and Yana Hasson and Eric Noland and Yuan Cao and Nathan Byrd and Le Hou and Qingze Wang and Thibault Sottiaux and Michela Paganini and Jean-Baptiste Lespiau and Alexandre Moufarek and Samer Hassan and Kaushik Shivakumar and Joost van Amersfoort and Amol Mandhane and Pratik Joshi and Anirudh Goyal and Matthew Tung and Andrew Brock and Hannah Sheahan and Vedant Misra and Cheng Li and Nemanja Rakićević and Mostafa Dehghani and Fangyu Liu and Sid Mittal and Junhyuk Oh and Seb Noury and Eren Sezener and Fantine Huot and Matthew Lamm and Nicola De Cao and Charlie Chen and Sidharth Mudgal and Romina Stella and Kevin Brooks and Gautam Vasudevan and Chenxi Liu and Mainak Chain and Nivedita Melinkeri and Aaron Cohen and Venus Wang and Kristie Seymore and Sergey Zubkov and Rahul Goel and Summer Yue and Sai Krishnakumaran and Brian Albert and Nate Hurley and Motoki Sano and Anhad Mohananey and Jonah Joughin and Egor Filonov and Tomasz Kępa and Yomna Eldawy and Jiawern Lim and Rahul Rishi and Shirin Badiezadegan and Taylor Bos and Jerry Chang and Sanil Jain and Sri Gayatri Sundara Padmanabhan and Subha Puttagunta and Kalpesh Krishna and Leslie Baker and Norbert Kalb and Vamsi Bedapudi and Adam Kurzrok and Shuntong Lei and Anthony Yu and Oren Litvin and Xiang Zhou and Zhichun Wu and Sam Sobell and Andrea Siciliano and Alan Papir and Robby Neale and Jonas Bragagnolo and Tej Toor and Tina Chen and Valentin Anklin and Feiran Wang and Richie Feng and Milad Gholami and Kevin Ling and Lijuan Liu and Jules Walter and Hamid Moghaddam and Arun Kishore and Jakub Adamek and Tyler Mercado and Jonathan Mallinson and Siddhinita Wandekar and Stephen Cagle and Eran Ofek and Guillermo Garrido and Clemens Lombriser and Maksim Mukha and Botu Sun and Hafeezul Rahman Mohammad and Josip Matak and Yadi Qian and Vikas Peswani and Pawel Janus and Quan Yuan and Leif Schelin and Oana David and Ankur Garg and Yifan He and Oleksii Duzhyi and Anton Älgmyr and Timothée Lottaz and Qi Li and Vikas Yadav and Luyao Xu and Alex Chinien and Rakesh Shivanna and Aleksandr Chuklin and Josie Li and Carrie Spadine and Travis Wolfe and Kareem Mohamed and Subhabrata Das and Zihang Dai and Kyle He and Daniel von Dincklage and Shyam Upadhyay and Akanksha Maurya and Luyan Chi and Sebastian Krause and Khalid Salama and Pam G Rabinovitch and Pavan Kumar Reddy M and Aarush Selvan and Mikhail Dektiarev and Golnaz Ghiasi and Erdem Guven and Himanshu Gupta and Boyi Liu and Deepak Sharma and Idan Heimlich Shtacher and Shachi Paul and Oscar Akerlund and François-Xavier Aubet and Terry Huang and Chen Zhu and Eric Zhu and Elico Teixeira and Matthew Fritze and Francesco Bertolini and Liana-Eleonora Marinescu and Martin Bölle and Dominik Paulus and Khyatti Gupta and Tejasi Latkar and Max Chang and Jason Sanders and Roopa Wilson and Xuewei Wu and Yi-Xuan Tan and Lam Nguyen Thiet and Tulsee Doshi and Sid Lall and Swaroop Mishra and Wanming Chen and Thang Luong and Seth Benjamin and Jasmine Lee and Ewa Andrejczuk and Dominik Rabiej and Vipul Ranjan and Krzysztof Styrc and Pengcheng Yin and Jon Simon and Malcolm Rose Harriott and Mudit Bansal and Alexei Robsky and Geoff Bacon and David Greene and Daniil Mirylenka and Chen Zhou and Obaid Sarvana and Abhimanyu Goyal and Samuel Andermatt and Patrick Siegler and Ben Horn and Assaf Israel and Francesco Pongetti and Chih-Wei "Louis" Chen and Marco Selvatici and Pedro Silva and Kathie Wang and Jackson Tolins and Kelvin Guu and Roey Yogev and Xiaochen Cai and Alessandro Agostini and Maulik Shah and Hung Nguyen and Noah Ó Donnaile and Sébastien Pereira and Linda Friso and Adam Stambler and Adam Kurzrok and Chenkai Kuang and Yan Romanikhin and Mark Geller and ZJ Yan and Kane Jang and Cheng-Chun Lee and Wojciech Fica and Eric Malmi and Qijun Tan and Dan Banica and Daniel Balle and Ryan Pham and Yanping Huang and Diana Avram and Hongzhi Shi and Jasjot Singh and Chris Hidey and Niharika Ahuja and Pranab Saxena and Dan Dooley and Srividya Pranavi Potharaju and Eileen O'Neill and Anand Gokulchandran and Ryan Foley and Kai Zhao and Mike Dusenberry and Yuan Liu and Pulkit Mehta and Ragha Kotikalapudi and Chalence Safranek-Shrader and Andrew Goodman and Joshua Kessinger and Eran Globen and Prateek Kolhar and Chris Gorgolewski and Ali Ibrahim and Yang Song and Ali Eichenbaum and Thomas Brovelli and Sahitya Potluri and Preethi Lahoti and Cip Baetu and Ali Ghorbani and Charles Chen and Andy Crawford and Shalini Pal and Mukund Sridhar and Petru Gurita and Asier Mujika and Igor Petrovski and Pierre-Louis Cedoz and Chenmei Li and Shiyuan Chen and Niccolò Dal Santo and Siddharth Goyal and Jitesh Punjabi and Karthik Kappaganthu and Chester Kwak and Pallavi LV and Sarmishta Velury and Himadri Choudhury and Jamie Hall and Premal Shah and Ricardo Figueira and Matt Thomas and Minjie Lu and Ting Zhou and Chintu Kumar and Thomas Jurdi and Sharat Chikkerur and Yenai Ma and Adams Yu and Soo Kwak and Victor Ähdel and Sujeevan Rajayogam and Travis Choma and Fei Liu and Aditya Barua and Colin Ji and Ji Ho Park and Vincent Hellendoorn and Alex Bailey and Taylan Bilal and Huanjie Zhou and Mehrdad Khatir and Charles Sutton and Wojciech Rzadkowski and Fiona Macintosh and Roopali Vij and Konstantin Shagin and Paul Medina and Chen Liang and Jinjing Zhou and Pararth Shah and Yingying Bi and Attila Dankovics and Shipra Banga and Sabine Lehmann and Marissa Bredesen and Zifan Lin and John Eric Hoffmann and Jonathan Lai and Raynald Chung and Kai Yang and Nihal Balani and Arthur Bražinskas and Andrei Sozanschi and Matthew Hayes and Héctor Fernández Alcalde and Peter Makarov and Will Chen and Antonio Stella and Liselotte Snijders and Michael Mandl and Ante Kärrman and Paweł Nowak and Xinyi Wu and Alex Dyck and Krishnan Vaidyanathan and Raghavender R and Jessica Mallet and Mitch Rudominer and Eric Johnston and Sushil Mittal and Akhil Udathu and Janara Christensen and Vishal Verma and Zach Irving and Andreas Santucci and Gamaleldin Elsayed and Elnaz Davoodi and Marin Georgiev and Ian Tenney and Nan Hua and Geoffrey Cideron and Edouard Leurent and Mahmoud Alnahlawi and Ionut Georgescu and Nan Wei and Ivy Zheng and Dylan Scandinaro and Heinrich Jiang and Jasper Snoek and Mukund Sundararajan and Xuezhi Wang and Zack Ontiveros and Itay Karo and Jeremy Cole and Vinu Rajashekhar and Lara Tumeh and Eyal Ben-David and Rishub Jain and Jonathan Uesato and Romina Datta and Oskar Bunyan and Shimu Wu and John Zhang and Piotr Stanczyk and Ye Zhang and David Steiner and Subhajit Naskar and Michael Azzam and Matthew Johnson and Adam Paszke and Chung-Cheng Chiu and Jaume Sanchez Elias and Afroz Mohiuddin and Faizan Muhammad and Jin Miao and Andrew Lee and Nino Vieillard and Jane Park and Jiageng Zhang and Jeff Stanway and Drew Garmon and Abhijit Karmarkar and Zhe Dong and Jong Lee and Aviral Kumar and Luowei Zhou and Jonathan Evens and William Isaac and Geoffrey Irving and Edward Loper and Michael Fink and Isha Arkatkar and Nanxin Chen and Izhak Shafran and Ivan Petrychenko and Zhe Chen and Johnson Jia and Anselm Levskaya and Zhenkai Zhu and Peter Grabowski and Yu Mao and Alberto Magni and Kaisheng Yao and Javier Snaider and Norman Casagrande and Evan Palmer and Paul Suganthan and Alfonso Castaño and Irene Giannoumis and Wooyeol Kim and Mikołaj Rybiński and Ashwin Sreevatsa and Jennifer Prendki and David Soergel and Adrian Goedeckemeyer and Willi Gierke and Mohsen Jafari and Meenu Gaba and Jeremy Wiesner and Diana Gage Wright and Yawen Wei and Harsha Vashisht and Yana Kulizhskaya and Jay Hoover and Maigo Le and Lu Li and Chimezie Iwuanyanwu and Lu Liu and Kevin Ramirez and Andrey Khorlin and Albert Cui and Tian LIN and Marcus Wu and Ricardo Aguilar and Keith Pallo and Abhishek Chakladar and Ginger Perng and Elena Allica Abellan and Mingyang Zhang and Ishita Dasgupta and Nate Kushman and Ivo Penchev and Alena Repina and Xihui Wu and Tom van der Weide and Priya Ponnapalli and Caroline Kaplan and Jiri Simsa and Shuangfeng Li and Olivier Dousse and Fan Yang and Jeff Piper and Nathan Ie and Rama Pasumarthi and Nathan Lintz and Anitha Vijayakumar and Daniel Andor and Pedro Valenzuela and Minnie Lui and Cosmin Paduraru and Daiyi Peng and Katherine Lee and Shuyuan Zhang and Somer Greene and Duc Dung Nguyen and Paula Kurylowicz and Cassidy Hardin and Lucas Dixon and Lili Janzer and Kiam Choo and Ziqiang Feng and Biao Zhang and Achintya Singhal and Dayou Du and Dan McKinnon and Natasha Antropova and Tolga Bolukbasi and Orgad Keller and David Reid and Daniel Finchelstein and Maria Abi Raad and Remi Crocker and Peter Hawkins and Robert Dadashi and Colin Gaffney and Ken Franko and Anna Bulanova and Rémi Leblond and Shirley Chung and Harry Askham and Luis C. Cobo and Kelvin Xu and Felix Fischer and Jun Xu and Christina Sorokin and Chris Alberti and Chu-Cheng Lin and Colin Evans and Alek Dimitriev and Hannah Forbes and Dylan Banarse and Zora Tung and Mark Omernick and Colton Bishop and Rachel Sterneck and Rohan Jain and Jiawei Xia and Ehsan Amid and Francesco Piccinno and Xingyu Wang and Praseem Banzal and Daniel J. Mankowitz and Alex Polozov and Victoria Krakovna and Sasha Brown and MohammadHossein Bateni and Dennis Duan and Vlad Firoiu and Meghana Thotakuri and Tom Natan and Matthieu Geist and Ser tan Girgin and Hui Li and Jiayu Ye and Ofir Roval and Reiko Tojo and Michael Kwong and James Lee-Thorp and Christopher Yew and Danila Sinopalnikov and Sabela Ramos and John Mellor and Abhishek Sharma and Kathy Wu and David Miller and Nicolas Sonnerat and Denis Vnukov and Rory Greig and Jennifer Beattie and Emily Caveness and Libin Bai and Julian Eisenschlos and Alex Korchemniy and Tomy Tsai and Mimi Jasarevic and Weize Kong and Phuong Dao and Zeyu Zheng and Frederick Liu and Fan Yang and Rui Zhu and Tian Huey Teh and Jason Sanmiya and Evgeny Gladchenko and Nejc Trdin and Daniel Toyama and Evan Rosen and Sasan Tavakkol and Linting Xue and Chen Elkind and Oliver Woodman and John Carpenter and George Papamakarios and Rupert Kemp and Sushant Kafle and Tanya Grunina and Rishika Sinha and Alice Talbert and Diane Wu and Denese Owusu-Afriyie and Cosmo Du and Chloe Thornton and Jordi Pont-Tuset and Pradyumna Narayana and Jing Li and Saaber Fatehi and John Wieting and Omar Ajmeri and Benigno Uria and Yeongil Ko and Laura Knight and Amélie Héliou and Ning Niu and Shane Gu and Chenxi Pang and Yeqing Li and Nir Levine and Ariel Stolovich and Rebeca Santamaria-Fernandez and Sonam Goenka and Wenny Yustalim and Robin Strudel and Ali Elqursh and Charlie Deck and Hyo Lee and Zonglin Li and Kyle Levin and Raphael Hoffmann and Dan Holtmann-Rice and Olivier Bachem and Sho Arora and Christy Koh and Soheil Hassas Yeganeh and Siim Põder and Mukarram Tariq and Yanhua Sun and Lucian Ionita and Mojtaba Seyedhosseini and Pouya Tafti and Zhiyu Liu and Anmol Gulati and Jasmine Liu and Xinyu Ye and Bart Chrzaszcz and Lily Wang and Nikhil Sethi and Tianrun Li and Ben Brown and Shreya Singh and Wei Fan and Aaron Parisi and Joe Stanton and Vinod Koverkathu and Christopher A. Choquette-Choo and Yunjie Li and TJ Lu and Abe Ittycheriah and Prakash Shroff and Mani Varadarajan and Sanaz Bahargam and Rob Willoughby and David Gaddy and Guillaume Desjardins and Marco Cornero and Brona Robenek and Bhavishya Mittal and Ben Albrecht and Ashish Shenoy and Fedor Moiseev and Henrik Jacobsson and Alireza Ghaffarkhah and Morgane Rivière and Alanna Walton and Clément Crepy and Alicia Parrish and Zongwei Zhou and Clement Farabet and Carey Radebaugh and Praveen Srinivasan and Claudia van der Salm and Andreas Fidjeland and Salvatore Scellato and Eri Latorre-Chimoto and Hanna Klimczak-Plucińska and David Bridson and Dario de Cesare and Tom Hudson and Piermaria Mendolicchio and Lexi Walker and Alex Morris and Matthew Mauger and Alexey Guseynov and Alison Reid and Seth Odoom and Lucia Loher and Victor Cotruta and Madhavi Yenugula and Dominik Grewe and Anastasia Petrushkina and Tom Duerig and Antonio Sanchez and Steve Yadlowsky and Amy Shen and Amir Globerson and Lynette Webb and Sahil Dua and Dong Li and Surya Bhupatiraju and Dan Hurt and Haroon Qureshi and Ananth Agarwal and Tomer Shani and Matan Eyal and Anuj Khare and Shreyas Rammohan Belle and Lei Wang and Chetan Tekur and Mihir Sanjay Kale and Jinliang Wei and Ruoxin Sang and Brennan Saeta and Tyler Liechty and Yi Sun and Yao Zhao and Stephan Lee and Pandu Nayak and Doug Fritz and Manish Reddy Vuyyuru and John Aslanides and Nidhi Vyas and Martin Wicke and Xiao Ma and Evgenii Eltyshev and Nina Martin and Hardie Cate and James Manyika and Keyvan Amiri and Yelin Kim and Xi Xiong and Kai Kang and Florian Luisier and Nilesh Tripuraneni and David Madras and Mandy Guo and Austin Waters and Oliver Wang and Joshua Ainslie and Jason Baldridge and Han Zhang and Garima Pruthi and Jakob Bauer and Feng Yang and Riham Mansour and Jason Gelman and Yang Xu and George Polovets and Ji Liu and Honglong Cai and Warren Chen and XiangHai Sheng and Emily Xue and Sherjil Ozair and Christof Angermueller and Xiaowei Li and Anoop Sinha and Weiren Wang and Julia Wiesinger and Emmanouil Koukoumidis and Yuan Tian and Anand Iyer and Madhu Gurumurthy and Mark Goldenson and Parashar Shah and MK Blake and Hongkun Yu and Anthony Urbanowicz and Jennimaria Palomaki and Chrisantha Fernando and Ken Durden and Harsh Mehta and Nikola Momchev and Elahe Rahimtoroghi and Maria Georgaki and Amit Raul and Sebastian Ruder and Morgan Redshaw and Jinhyuk Lee and Denny Zhou and Komal Jalan and Dinghua Li and Blake Hechtman and Parker Schuh and Milad Nasr and Kieran Milan and Vladimir Mikulik and Juliana Franco and Tim Green and Nam Nguyen and Joe Kelley and Aroma Mahendru and Andrea Hu and Joshua Howland and Ben Vargas and Jeffrey Hui and Kshitij Bansal and Vikram Rao and Rakesh Ghiya and Emma Wang and Ke Ye and Jean Michel Sarr and Melanie Moranski Preston and Madeleine Elish and Steve Li and Aakash Kaku and Jigar Gupta and Ice Pasupat and Da-Cheng Juan and Milan Someswar and Tejvi M. and Xinyun Chen and Aida Amini and Alex Fabrikant and Eric Chu and Xuanyi Dong and Amruta Muthal and Senaka Buthpitiya and Sarthak Jauhari and Nan Hua and Urvashi Khandelwal and Ayal Hitron and Jie Ren and Larissa Rinaldi and Shahar Drath and Avigail Dabush and Nan-Jiang Jiang and Harshal Godhia and Uli Sachs and Anthony Chen and Yicheng Fan and Hagai Taitelbaum and Hila Noga and Zhuyun Dai and James Wang and Chen Liang and Jenny Hamer and Chun-Sung Ferng and Chenel Elkind and Aviel Atias and Paulina Lee and Vít Listík and Mathias Carlen and Jan van de Kerkhof and Marcin Pikus and Krunoslav Zaher and Paul Müller and Sasha Zykova and Richard Stefanec and Vitaly Gatsko and Christoph Hirnschall and Ashwin Sethi and Xingyu Federico Xu and Chetan Ahuja and Beth Tsai and Anca Stefanoiu and Bo Feng and Keshav Dhandhania and Manish Katyal and Akshay Gupta and Atharva Parulekar and Divya Pitta and Jing Zhao and Vivaan Bhatia and Yashodha Bhavnani and Omar Alhadlaq and Xiaolin Li and Peter Danenberg and Dennis Tu and Alex Pine and Vera Filippova and Abhipso Ghosh and Ben Limonchik and Bhargava Urala and Chaitanya Krishna Lanka and Derik Clive and Yi Sun and Edward Li and Hao Wu and Kevin Hongtongsak and Ianna Li and Kalind Thakkar and Kuanysh Omarov and Kushal Majmundar and Michael Alverson and Michael Kucharski and Mohak Patel and Mudit Jain and Maksim Zabelin and Paolo Pelagatti and Rohan Kohli and Saurabh Kumar and Joseph Kim and Swetha Sankar and Vineet Shah and Lakshmi Ramachandruni and Xiangkai Zeng and Ben Bariach and Laura Weidinger and Tu Vu and Alek Andreev and Antoine He and Kevin Hui and Sheleem Kashem and Amar Subramanya and Sissie Hsiao and Demis Hassabis and Koray Kavukcuoglu and Adam Sadovsky and Quoc Le and Trevor Strohman and Yonghui Wu and Slav Petrov and Jeffrey Dean and Oriol Vinyals},
      year={2025},
      eprint={2312.11805},
      archivePrefix={arXiv},
      primaryClass={cs.CL},
      url={https://arxiv.org/abs/2312.11805}, 
}

@article{cowan2001magical,
  title={The magical number 4 in short-term memory: A reconsideration of mental storage capacity},
  author={Cowan, Nelson},
  journal={Behavioral and Brain Sciences},
  volume={24},
  number={1},
  pages={87--114},
  year={2001},
  publisher={Cambridge University Press}
}

@misc{shan2025cognitivememorylargelanguage,
      title={Cognitive Memory in Large Language Models}, 
      author={Lianlei Shan and Shixian Luo and Zezhou Zhu and Yu Yuan and Yong Wu},
      year={2025},
      eprint={2504.02441},
      archivePrefix={arXiv},
      primaryClass={cs.CL},
      url={https://arxiv.org/abs/2504.02441}, 
}

@inproceedings{
tavakoli2026beyond,
title={Beyond a Million Tokens: Benchmarking and Enhancing Long-Term Memory in {LLM}s},
author={Mohammad Tavakoli and Alireza Salemi and Carrie Ye and Mohamed Abdalla and Hamed Zamani and J Ross Mitchell},
booktitle={The Fourteenth International Conference on Learning Representations (ICLR'26)},
year={2026},
url={https://openreview.net/forum?id=y59hf5lrMn}
}

@article{werfel2026fluid,
  title={Fluid thinking about collective intelligence},
  author={Werfel, Justin},
  journal={Nature Machine Intelligence},
  volume={8},
  number={4},
  pages={506--516},
  year={2026},
  publisher={Nature Publishing Group UK London}
}

@article{zhu2024self,
  title={Self-organizing nervous systems for robot swarms},
  author={Zhu, Weixu and O{\u{g}}uz, Sinan and Heinrich, Mary Katherine and Allwright, Michael and Wahby, Mostafa and Christensen, Anders Lyhne and Garone, Emanuele and Dorigo, Marco},
  journal={Science Robotics},
  volume={9},
  number={96},
  pages={eadl5161},
  year={2024},
  publisher={American Association for the Advancement of Science}
}

@article{david2009trail,
  title={Trail pheromones of ants},
  author={David Morgan, E},
  journal={Physiological Entomology},
  volume={34},
  number={1},
  pages={1--17},
  year={2009},
  publisher={Wiley Online Library}
}

@article{salman2024automatic,
  title={Automatic design of stigmergy-based behaviours for robot swarms},
  author={Salman, Muhammad and Garz{\'o}n Ramos, David and Birattari, Mauro},
  journal={Communications Engineering},
  volume={3},
  number={1},
  pages={30},
  year={2024},
  publisher={Nature Publishing Group UK London}
}

@ARTICLE{11124525,
  author={Jond, Hossein B.},
  journal={IEEE Transactions on Cognitive and Developmental Systems}, 
  title={A Minimal Model for Emergent Collective Behaviors in Autonomous Robotic Multiagent Systems}, 
  year={2026},
  volume={18},
  number={2},
  pages={373-384},
  doi={10.1109/TCDS.2025.3598690}}

@article{hassan2025agentic,
  title={Agentic software engineering: Foundational pillars and a research roadmap},
  author={Hassan, Ahmed E and Li, Hao and Lin, Dayi and Adams, Bram and Chen, Tse-Hsun and Kashiwa, Yutaro and Qiu, Dong},
  journal={arXiv preprint arXiv:2509.06216},
  year={2025}
}

@ARTICLE{1663532,
  author={Brooks},
  journal={Computer}, 
  title={No silver bullet--Essence and accidents of software engineering}, 
  year={1987},
  volume={20},
  number={4},
  pages={10-19},
  doi={10.1109/MC.1987.1663532}}

@inbook{Newell1995,
author = {Newell, Allen and Simon, H. A.},
title = {GPS, a program that simulates human thought},
year = {1995},
isbn = {0262560925},
publisher = {MIT Press},
address = {Cambridge, MA, USA},
booktitle = {Computers \& Thought},
pages = {279–293},
numpages = {15}
}

@book{minsky1986society,
  title={Society of Mind},
  author={Minsky, Marvin},
  year={1986},
  publisher={Simon and Schuster}
}

@article{rinta2023getting,
  title={Getting trapped in technical debt: Sociotechnical analysis of a legacy system’s replacement},
  author={Rinta-Kahila, Tapani and Penttinen, Esko and Lyytinen, Kalle},
  journal={MIS Quarterly},
  volume={47},
  number={1},
  pages={1--32},
  year={2023},
  publisher={Management Information Systems Research Center, University of Minnesota}
}

@techreport{hong2025context,
  title = {Context Rot: How Increasing Input Tokens Impacts LLM Performance},
  author = {Hong, Kelly and Troynikov, Anton and Huber, Jeff},
  year = {2025},
  month = {July},
  institution = {Chroma},
  url = {https://trychroma.com/research/context-rot},
}

@misc{yang2026programbenchlanguagemodelsrebuild,
      title={ProgramBench: Can Language Models Rebuild Programs From Scratch?}, 
      author={John Yang and Kilian Lieret and Jeffrey Ma and Parth Thakkar and Dmitrii Pedchenko and Sten Sootla and Emily McMilin and Pengcheng Yin and Rui Hou and Gabriel Synnaeve and Diyi Yang and Ofir Press},
      year={2026},
      eprint={2605.03546},
      archivePrefix={arXiv},
      primaryClass={cs.SE},
      url={https://arxiv.org/abs/2605.03546}, 
}

@misc{wang2026humansmissingaicoding,
      title={Humans are Missing from AI Coding Agent Research}, 
      author={Zora Z. Wang and John Yang and Kilian Lieret and Alexa Tartaglini and Valerie Chen and Yuxiang Wei and Zijian Wang and Lingming Zhang and Karthik Narasimhan and Ludwig Schmidt and Graham Neubig and Daniel Fried and Diyi Yang},
      year={2026},
      eprint={2608.12355},
      archivePrefix={arXiv},
      primaryClass={cs.HC},
      url={https://arxiv.org/abs/2608.12355}, 
}

@article{welsh2023end,
  title={The end of programming},
  author={Welsh, Matt},
  journal={Communications of the ACM},
  volume={66},
  number={1},
  pages={34--35},
  year={2023},
  publisher={ACM New York, NY, USA}
}

\section*{Funding}
This work is partially supported by National Natural Science Foundation under the grant 62076232. 

\section*{Code availability}
No code was used to compute the results. The example codes are shown in Fig.~\ref{fig:fig2}.

\section*{Author contributions}
YW, ZJ, and GZ conceptualized the Perspective. ZJ prepared the visualizations. YW, ZJ, MZ, and YF drafted the original manuscript. YW, GZ, and YF reviewed and edited the manuscript.

\section*{Ethics declarations}
\subsection*{Competing interests.}
We declare no competing financial interests or personal relationships that could have influenced the work reported in this study. The manuscript's publication is approved by all authors and explicitly by the responsible authorities where the work was carried out.

\end{document}